\documentclass[a4paper,12pt]{article}

\usepackage{amstext,amsmath,amssymb,amsthm}  % AMS Math
\usepackage{txfonts}  % Public Times New Roman text & math font
\usepackage{graphicx}  % For figures

\usepackage[square,comma,numbers,sort&compress]{natbib}
\usepackage{hypernat}  % For natbib to interact with hyperref

\usepackage{comment}  % For multiline comments

\usepackage{authblk}

\begin{document}

\numberwithin{equation}{section}
\numberwithin{figure}{section}

\allowdisplaybreaks[1]  % It allows multiple-lines equations to broken

\title{Exact and efficient calculation of Lagrange multipliers in constrained biological polymers: Proteins and nucleic acids as example cases}

\author[,2,1,3]{Pablo Garc\'{\i}a-Risue\~no\footnote{Email: {\tt garcia.risueno@gmail.com}}}
\author[2,1,3]{Pablo Echenique}
\author[3,1]{J. L. Alonso}

\affil[1]{Instituto de Biocomputaci\'on y F{\'{\i}}sica de Sistemas Complejos 
(BIFI), Universidad de Zaragoza, Mariano Esquillor s/n, Edificio I+D, E-50018 
Zaragoza, Spain}
\affil[2]{Instituto de Qu\'{\i}mica F\'{\i}sica Rocasolano, CSIC, Serrano 
119, E-28006 Madrid, Spain}
\affil[3]{Departamento de F{\'{\i}}sica Te\'orica, Universidad de Zaragoza, 
Pedro Cerbuna 12, E-50009 Zaragoza, Spain}

\date{\today}

\maketitle

% With an abstract
\begin{abstract}
In order to accelerate molecular dynamics simulations, it is very common to
impose holonomic constraints on their hardest degrees of freedom. In this way,
the time step used to integrate the equations of motion can be increased, thus
allowing, in principle, to reach longer total simulation times. The imposition
of such constraints results in an aditional set of $N_c$ equations (the
equations of constraint) and unknowns (their associated Lagrange multipliers),
that must be solved in one way or another at each time step of the dynamics.
In this work it is shown that, due to the essentially linear structure of
typical biological polymers, such as nucleic acids or proteins, the algebraic
equations that need to be solved involve a matrix which is banded if the
constraints are indexed in a clever way. This allows to obtain the Lagrange
multipliers through a non-iterative procedure, which can be considered exact
up to machine precision, and which takes $O(N_c)$ operations, instead of the
usual $O(N_c^3)$ for generic molecular systems. We develop the formalism, and
describe the appropriate indexing for a number of model molecules and also for
alkanes, proteins and DNA. Finally, we provide a numerical example of the
technique in a series of polyalanine peptides of different lengths using the
AMBER molecular dynamics package.
\vspace{0.4cm}\\ {\bf Keywords:} constraints, Lagrange multipliers, banded systems, molecular dynamics, proteins, DNA
\vspace{0.2cm}\\

\end{abstract}

\section{Introduction}
\label{sec:introduction}

Due to the high frequency of the fastest internal motions in molecular
systems, the discrete time step for molecular dynamics simulations must be
very small (of the order of femtoseconds), while the actual span of
biochemical proceses typically require the choice of relatively long total
times for simulations (e.g., from microseconds to milliseconds for protein
folding processes). In addition to this, since biologically interesting
molecules (like proteins \cite{Ech2007COP} and DNA \cite{Ced2009NRG}) 
consist of thousands of atoms, their trajectories in
configuration space are esentially chaotic, and therefore reliable quantities
can be obtained from the simulation only after statistical analysis
\cite{Pia2011JMB}. In order to cope with these two requirements, which force
the computation of a large number of dynamical steps if predictions want to be
made, great efforts are being done both in hardware
\cite{Sha2009XXX,dJo2010PCCP} and in software \cite{Phi2005JCC,Pea1995CoPCo}
solutions. In fact, only in very recent times, simulations for interesting
systems of hundreds of thousand of atoms in the millisecond scale are starting
to become affordable, being still, as we mentioned, the main limitation of
these computational techniques the large difference between the elemental time
step used to integrate the equations of motion and the total time span needed
to obtain useful information. In this context, strategies to increase the time
step are very valuable.

A widely used method to this end is to constrain some of the internal degrees
of freedom \cite{Gon2009JCP} of a molecule (typically bond lengths, sometimes bond
angles and rarely dihedral angles. For a Verlet-like integrator
\cite{Ver1967PR,Fre2002BOOK}, stability requires the time step to be at least
about five times smaller than the period of the fastest vibration in the
studied system \cite{Fee1999JCC}. Here is where constraints come into play. By
constraining the hardest degrees of freedom, the fastest vibrational motions
are frozen, and thus larger time steps still produce stable simulations. If
constraints are imposed on bond lengths involving hydrogens, the time step can
typically be increased by a factor of 2 to 3 (from 1 fs to 2 or 3 fs)
\cite{Eas2010JCTC}. Constraining additional internal degrees of freedom, such
as heavy atoms bond lengths and bond angles, allows even larger timesteps
\cite{Maz1998jpcb,Fee1999JCC}, but one has to be careful, since, as more and
softer degrees of freedom are constrained, the more likely it is that the
physical properties of the simulated system could be severely distorted
\cite{Ech2006JCC2,VGu1982MM,Bar1995JCoP}.

The essential ingredient in the calculation of the forces produced by the
imposition of constraints are the so-called Lagrange multipliers
\cite{Gol2002BOOK}, and their efficient numerical evaluation is therefore of
the utmost importance. In this work, we show that the fact that many
interesting biological molecules are esentially linear polymers allows to
calculate the Lagrange multipliers in order $N_c$ operations (for a molecule
where $N_c$ constraints are imposed) in an exact (up to machine precision),
non-iterative way. Moreover, we provide a method to do so which is based in a
clever ordering of the constraints indices, and in a recently introduced
algorithm for solving linear banded systems \cite{GR2010JCoP}. It is worth
mentioning that, in the specialized literature, this possibility has not been
considered as far as we are aware; with some works commenting that solving
this kind of linear problems (or related ones) is costly (but not giving
further details) \cite{Ryc1977JCOP,Dil1987JCC,Cic1986CPR}, and some other
works explicitly stating that such a computation must take $O(N_c^3)$
 \cite{Kra2000JCC} or $O(N_c^2)$ \cite{Gon2006JCP,Maz2007JPA} operations. 
 Also, in the field of robot kinematics, many
$O(N_c)$ algorithms have been devised to deal with different aspects of
constrained physical systems (robots in this case) \cite{Fea1999IJRR,Bae1987MSM1,Lee2007Rob},
but none of them tackles the calculation of the Lagrange multipliers themselves.

This work is structured as follows. In sec.~\ref{sec_aclm}, we introduce the
basic formalism for the calculation of constraint forces and Lagrange
multipliers. In sec.~\ref{soc}, we explain how to index the constraints in
order for the resulting linear system of equations to be banded with the
minimal bandwidth (which is essential to solve it efficiently). We do this
starting by very simple toy systems and building on complexity as we move
forward towards the final discussion about DNA and proteins; this way of
proceeding is intended to help the reader build the corresponding indexing for
molecules not covered in this work. In sec.~\ref{sec:numerical}, we apply the
introduced technique to a polyalanine peptide using the AMBER molecular dynamics package and
we compare the relative efficiency between the calculation of the Lagrange
multipliers in the traditional way ($O(N_c^3)$) and in the new way presented
here ($O(N_c)$). Finally, in sec.~\ref{sec:conclusions}, we summarize the main
conclusions of this work and outline some possible future applications.

\section{Calculation of the Lagrange multipliers}
\label{sec_aclm}

If holonomic, rheonomous constraints are imposed on a classical system of $n$
atoms, and the D'Alem\-bert's principle is assumed to hold, its motion is the
solution of the following system of differential equations
\cite{Gol2002BOOK,Hen1992BOOK}:
\begin{subequations}
\label{sistemBasico}
\begin{align}
m_\alpha \frac{\mathrm{d}^{2}\vec{x}_{\alpha}(t)}{\mathrm{d}t^{2}}
 & = \vec{F}_{\alpha}(x(t))+
 \sum_{I=1}^{N_{c}}{\lambda_{I}(t)\vec{\nabla}_{\alpha} \sigma^{I}(x(t))} \ ,
  \qquad \alpha=1,\ldots,n \ , \label{newton} \\
\sigma^{I}(x(t)) & =0 \ , \qquad I=1,\ldots,{N_{c}} \ , \label{constr} \\
x(t_{0}) & = x_{0}\ , \label{sistemaBasico_ci1} \\
\frac{\mathrm{d}x(t_{0})}{\mathrm{d}t} & =\dot{x}_{0} \label{sistemaBasico_ci2} \ ,
\end{align}
\end{subequations}
where~(\ref{newton}) is the modified Newton's second law and~(\ref{constr})
are the equations of the constraints themselves; $\lambda_I$ are the Lagrange
multipliers associated with the constraints; $\vec{F}_{\alpha}$ represents the
external force acting on atom $\alpha$, $\vec{x}_{\alpha}$ is its Euclidean
position, and $x$ colectively denote the set of all such coordinates. We
assume $\vec{F}_{\alpha}$ to be conservative, i.e., to come from the gradient
of a scalar potential function $V(x)$; and
$\sum_{I=1}^{{N_{c}}}\lambda_{I}\vec{\nabla}_{\alpha} \sigma^{I}(x)$ should be
regarded as the \emph{force of constraint} acting on atom $\alpha$.

Also, in the above expression and in this whole document we will use the
following notation for the different indices:

\begin{itemize}

\item $\alpha,\beta,\gamma.\epsilon,\zeta=1,\ldots,n$ (except if otherwise 
stated) for atoms.

\item $\mu,\nu=1,\ldots,3n$ (except if otherwise stated) for the atoms
coordinates when no explicit reference to the atom index needs to be made.

\item $I,J=1,\dots,N_c$ for constrains and the rows and columns of the 
associated matrices.

\item $k,l$ as generic indices for products and sums.

\end{itemize}

The existence of $N_{c}$ constraints turns a system of $N=3n$ differential
equations with $N$ unknowns into a system of $N+{N_{c}}$
algebraic-differential equations with $N+{N_{c}}$ unknowns. The constraints
equations in~(\ref{constr}) are the new equations, and the Lagrange
multipliers are the new unknowns whose value must be found in order to solve
the system.

If the functions $\sigma^I(x)$ are analytical, the system of equations
in~(\ref{sistemBasico}) is equivalent to the following one:
\begin{subequations}
\begin{align}
m_{\alpha}\frac{\mathrm{d}^{2}\vec{x}_{\alpha}(t)}{\mathrm{d}t^{2}}
 & = \vec{F}_{\alpha}(x(t)) + 
  \sum_{I=1}^{{N_{c}}}{\lambda_{I}(t)\vec{\nabla}_{\alpha} 
  \sigma^{I}(x(t))} \ , \label{newton2} \\ 
\sigma^{I}(x(t_{0})) & = 0 \ , \\
\frac{\mathrm{d}\sigma^{I}(x(t_{0}))}{\mathrm{d}t} & = 0 \ , \\
\frac{\mathrm{d}^{2}\sigma^{I}(x(t))}{\mathrm{d}t^{2}} & = 0 \ ,
   \quad \forall t \  , \label{constr2} \\ 
x(t_{0}) & = x_{0} \ , \\
\frac{\mathrm{d}x(t_{0})}{\mathrm{d}t} & = \dot{x}_{0} \ .
\end{align}
\end{subequations}

In this new form, it exists a more direct path to solve for the Lagrange
multipliers: If we explicitly calculate the second derivative in
eq.~(\ref{constr2}) and then substitute eq.~(\ref{newton2}) where the
accelerations appear, we arrive to
\begin{eqnarray}
\label{Rl}
\frac{\mathrm{d}^{2}\sigma^{I}}{\mathrm{d}t^{2}}&=&
\sum_{\mu} \frac{1}{m_{\mu}} \left(F_{\mu}+
\sum_{J}\lambda_{J}\frac{\partial \sigma^{J}}{\partial 
 x^{\mu}}\right)\frac{\partial \sigma^{I}}{\partial x^{\mu}}+
 \sum_{\mu,\nu}\frac{\mathrm{d} x^{\mu}}{\mathrm{d}t}
 \frac{\mathrm{d} x^{\nu}}{\mathrm{d}t}\frac{\partial^{2}
  \sigma^{I}}{\partial x^{\mu} \partial x^{\nu}} \nonumber \\
 & =: &p^{I}+q^{I}+\sum_{J}R_{IJ}\lambda_{J}=0 \ , \qquad I=1,\ldots,N_c \ ,
\end{eqnarray}
where we have implicitly defined
\begin{subequations}
\label{PQ}
\begin{align}
p^{I} & :=\sum_{\mu}
  \frac{1}{m_{\mu}}F_{\mu}\frac{\partial \sigma^{I}}{\partial x_{\mu}}=
  \sum_{\alpha} \frac{1}{m_{\alpha}}
  \vec{F}_{\alpha} \cdot \vec{\nabla}_{\alpha}\sigma^{I} \ ,
  \label{PQ1} \\
q^{I} & := \sum_{\mu,\nu}\frac{\mathrm{d} x^{\mu}}{\mathrm{d}t}\frac{\mathrm{d}x^{\nu}}{\mathrm{d}t}
  \frac{\partial^{2} \sigma^{I}}{\partial x^{\mu} \partial x^{\nu}} \ ,
  \label{PQ2} \\
R_{IJ} & := \sum_{\mu} \frac{1}{m_{\mu}}\frac{\partial \sigma^{I}}{\partial
  x^{\mu}}\frac{\partial \sigma^{J}}{\partial x^{\mu}}= 
  \sum_{\alpha}{\frac{1}{m_{\alpha}}\vec{\nabla}_{\alpha}\sigma^{I}\cdot
  \vec{\nabla}_{\alpha}\sigma^{J}} \ , \label{defR}
\end{align}
\end{subequations}
and it becomes clear that, at each $t$, the Lagrange multipliers $\lambda_J$
are actually a \emph{known} function of the positions and the velocities.

We shall use the shorthand
\begin{equation}
\label{defO}
o^{I}:=p^{I}+q^{I} \ , \qquad I=1,\ldots, N_{c} \ ,
\end{equation}
and, $o$, $p$, and $q$ to denote the whole $N_c$-tuples, as usual.

Now, in order to obtain the Lagrange multipliers $\lambda_{J}$, we just need
to solve
\begin{equation}
\label{lm}
\sum_{I}R_{IJ}\lambda_J = - \left(p^{I}+q^{I}\right) \quad 
\Rightarrow \quad R\lambda = -o \ .
\end{equation}

This is a linear system of $N_{c}$ equations and $N_{c}$ unknowns. In the
following, we will prove that the solution to it, when constraints are
imposed on typical biological polymers, can be found in $O(N_{c})$ operations
without the use of any iterative or truncation procedure, i.e., in an exact
way up to machine precision. To show this, first, we will prove that the value
of the vectors $p$ and $q$ can be obtained in $O(N_{c})$ operations. Then, we
will show that the same is true for all the non-zero entries of matrix $R$,
and finally we will briefly discuss the results in \cite{GR2010JCoP}, where we
introduced an algorithm to solve the system in~(\ref{lm}) also in $O(N_{c})$
operations.

It is worth remarking at this point that, in this work, we will only consider
constraints that hold the distance between pairs of atoms constant, i.e.,
\begin{equation}
\label{sigma_generica}
\sigma^{I(\alpha,\beta)}(x)
 :=|\vec{x_{\alpha}}-\vec{x}_{\beta}|^{2}-(a_{\alpha,\beta})^{2}\ ,
\end{equation}
where $a_{\alpha,\beta}$ is a constant number, and the fact that we can
establish a correspondence between constrained pairs ($\alpha,\beta$)
and the constraints indices has been explicitly indicated by the notation
$I(\alpha,\beta)$.

This can represent a constraint on:

\begin{itemize}

\item a bond length between atoms $\alpha$ and $\beta$,

\item a bond angle between atoms $\alpha$, $\beta$ and $\gamma$, if both
$\alpha$ and $\beta$ are connected to $\gamma$ through constrained bond
lengths,

\item a principal dihedral angle involving $\alpha$, $\beta$, $\gamma$ and
$\delta$ (see \cite{Ech2006JCCa} for a rigorous definition of the different
types of internal coordinates), if the bond lengths ($\alpha,\beta$),
($\beta,\gamma$) and ($\gamma,\delta$) are constrained, as well as the bond
angles ($\alpha,\beta,\gamma$) and ($\beta,\gamma,\delta$),

\item or a phase dihedral angle involving $\alpha$, $\beta$, $\gamma$ and
$\delta$ if the bond lengths ($\alpha,\beta$), ($\beta,\gamma$) and
($\beta,\delta$) are constrained, as well as the bond angles
($\alpha,\beta,\gamma$) and ($\alpha,\beta,\delta$).

\end{itemize}

This way to constrain degrees of freedom is called \emph{triangularization}.
If no triangularization is desired (as, for example, if we want to constrain
dihedral angles but not bond angles), different explicit expressions than
those in the following paragraphs must be written down, but the basic concepts
introduced here are equally valid and the main conclusions still hold.

Now, from eq.~(\ref{sigma_generica}), we obtain 
\begin{equation}
\label{grad_sigma}
\vec{\nabla}_{\gamma}\sigma^{I(\alpha,\beta)}=
2(\vec{x_{\alpha}}-\vec{x}_{\beta})(\delta_{\gamma,\alpha}-
 \delta_{\gamma,\beta}) \ .
\end{equation}

Inserting this into~(\ref{PQ1}), we get a simple expression for 
$p^{I(\alpha,\beta)}$
\begin{eqnarray}
\label{neoP}
p^{I(\alpha,\beta)} &:=& \sum_{\mu} \frac{1}{m_{\mu}}F_{\mu}\frac{\partial
  \sigma^{I(\alpha,\beta)}}{\partial x_{\mu}}=
 \sum_{\gamma}\frac{1}{m_{\gamma}}\vec{F}_{\gamma} \cdot 
 \vec{\nabla}_{\gamma}\sigma^{I(\alpha,\beta)} \\
 & = &\sum_{\gamma}\frac{2}{m_{\gamma}}\vec{F}_{\gamma} \cdot 
 (\vec{x_{\alpha}}-\vec{x}_{\beta})(\delta_{\gamma,\alpha}-
  \delta_{\gamma,\beta})=
 2(\vec{x_{\alpha}}-\vec{x}_{\beta}) \cdot \left(
 \frac{\vec{F}_{\alpha}}{m_{\alpha}}-\frac{\vec{F}_{\beta}}{m_{\beta}}\right)  
 \nonumber \ .
\end{eqnarray}

The calculation of $q^{I(\alpha,\beta)}$ is more involved, but it also results 
into a simple expression: First, we remember that the indices run as 
$\mu,\nu=1,\ldots,3n$, and $\alpha=1,\ldots,n$, and we produce the following
trivial relationship:
\begin{eqnarray}
\vec{x}_{\alpha} & = & 
 x^{3\alpha-2}\hat{i}+x^{3\alpha-1}\hat{j}+x^{3\alpha}\hat{k}
 \nonumber \\
\Rightarrow \quad \frac{\partial \vec{x}_{\alpha}}{\partial
x^{\mu}} & = & \frac{\partial
  (x^{3\alpha-2}\hat{i}+x^{3\alpha-1}\hat{j}+x^{3\alpha}\hat{k})}
  {\partial x^{\mu}} 
   =\delta_{3\alpha-2,\mu}\hat{i}
   +\delta_{3\alpha-1,\mu}\hat{j}
   +\delta_{3\alpha,\mu}\hat{k} \ ,
\end{eqnarray}
where $\hat{i}$, $\hat{j}$ and $\hat{k}$ are the unitary vectors along the
$x$, $y$ and $z$ axes, respectively.

Therefore, much related to eq.~(\ref{grad_sigma}), we can compute the first
derivative of $\sigma^{I(\alpha,\beta)}$:
\begin{eqnarray}
\frac{\partial \sigma^{I(\alpha,\beta)}}{\partial x^{\mu}} & = & 
  \frac{\partial 
    ((\vec{x}_{\alpha}-\vec{x}_{\beta})^{2}-a^{2}_{\alpha,\beta})}
  {\partial x^{\mu}} \nonumber \\
 & = & 2 (\vec{x}_{\alpha}-\vec{x}_{\beta}) \cdot 
  [(\delta_{3\alpha-2,\mu}\hat{i} +    
    \delta_{3\alpha-1,\mu}\hat{j} + 
    \delta_{3\alpha,\mu}\hat{k}) \nonumber \\
 & & \qquad \qquad \qquad
 - (\delta_{3\beta-2,\mu}\hat{i}  
    +\delta_{3\beta-1,\mu}\hat{j}
    +\delta_{3\beta,\mu}\hat{k})] \ ,
\end{eqnarray}
and also the second derivative:
\begin{eqnarray}
\frac{\partial^{2} \sigma^{I(\alpha,\beta)}}
     {\partial x^{\mu}\partial x^{\nu}} & = & 
2 [(\delta_{3\alpha-2,\mu}\hat{i}
   +\delta_{3\beta,\mu}\hat{j}
   +\delta_{3\alpha,\mu}\hat{k})
  -(\delta_{3\beta-2,\mu}\hat{i}
   +\delta_{3\beta-1,\mu}\hat{j}
   +\delta_{3\beta,\mu}\hat{k})] \nonumber \\
 & & \mbox{} \cdot 
 [(\delta_{3\alpha-2,\nu}\hat{i}
  +\delta_{3\alpha-1,\nu}\hat{j}
  +\delta_{3\alpha,\nu}\hat{k})
 -(\delta_{3\beta-2,\nu}\hat{i}
  +\delta_{3\beta-1,\nu}\hat{j}
  +\delta_{3\beta,\nu}\hat{k})] \nonumber \\
 & = & 2 
 ( \delta_{3\alpha-2,\mu}\delta_{3\alpha-2,\nu}
  +\delta_{3\beta-2,\mu}\delta_{3\beta-2,\nu}
  -\delta_{3\alpha-2,\mu}\delta_{3\beta-2,\nu}
  -\delta_{3\beta-2,\mu}\delta_{3\alpha-2,\nu} \nonumber \\
 & & \mbox{} 
  +\delta_{3\alpha-1,\mu}\delta_{3\alpha-1,\nu}
  +\delta_{3\beta-1,\mu}\delta_{3\beta-1,\nu}
  -\delta_{3\alpha-1,\mu}\delta_{3\beta-1,\nu}
  -\delta_{3\beta-1,\mu}\delta_{3\alpha-1,\nu} \nonumber \\
 & & \mbox{} 
  +\delta_{3\alpha,\mu}\delta_{3\alpha,\nu}
  +\delta_{3\beta,\mu}\delta_{3\beta,\nu}
  -\delta_{3\alpha,\mu}\delta_{3\beta,\nu}
  -\delta_{3\beta,\mu}\delta_{3\alpha,\nu}) \ .
\end{eqnarray}

Taking this into the original expression for $q^{I(\alpha,\beta)}$ in
eq.~(\ref{PQ2}) and playing with the sums and the deltas, we arrive to
\begin{eqnarray}
\label{neoQ}
q^{I(\alpha,\beta)} & := & \sum_{\mu,\nu}\frac{\mathrm{d} x^{\mu}}
      {\mathrm{d}t}
 \frac{\mathrm{d}x^{\nu}}{\mathrm{d}t}
 \frac{\partial^{2} \sigma^{I(\alpha,\beta)}}
 {\partial x^{\mu} \partial x^{\nu}} \nonumber \\
 & = &
  2 \left(\frac{\mathrm{d}x^{3\alpha-2}}{\mathrm{d}t} \right)^{2}
+ 2 \left(\frac{\mathrm{d}x^{3\beta-2}}{\mathrm{d}t} \right)^{2}
- 4 \left(\frac{\mathrm{d}x^{3\alpha-2}}{\mathrm{d}t}
    \frac{dx^{3\beta-2}}{dt} \right)
  \nonumber \\
 & & \mbox{}
+ 2\left(\frac{\mathrm{d}x^{3\alpha-1}}{\mathrm{d}t} \right)^{2}
+ 2\left(\frac{\mathrm{d}x^{3\beta-1}}{\mathrm{d}t}\right)^{2}
- 4\left(\frac{\mathrm{d}x^{3\alpha-1}}{\mathrm{d}t}
  \frac{\mathrm{d}x^{3\beta-1}}{\mathrm{d}t} \right)
  \nonumber \\
 & & \mbox{}
+ 2\left(\frac{\mathrm{d}x^{3\alpha}}{\mathrm{d}t} \right)^{2}
+ 2\left(\frac{\mathrm{d}x^{3\beta}}{\mathrm{d}t} \right)^{2}
- 4\left(\frac{\mathrm{d}x^{3\alpha}}{\mathrm{d}t}
  \frac{\mathrm{d}x^{3\beta}}{\mathrm{d}t}\right)
  \nonumber \\
 & = & 2\left|\frac{\mathrm{d}\vec{x}_{\alpha}}{\mathrm{d}t}
    -\frac{\mathrm{d}\vec{x}_{\beta}}{\mathrm{d}t} \right|^{2} \ .
\end{eqnarray}

Now, eqs.~(\ref{defO}), (\ref{neoP}) and (\ref{neoQ}) can be gathered together 
to become
\begin{equation}
\label{termindeporden1}
o^{I(\alpha,\beta)}= 2\left|\frac{\mathrm{d}\vec{x}_{\alpha}}{\mathrm{d}t}
-\frac{\mathrm{d}\vec{x}_{\beta}}{\mathrm{d}t} \right|^{2}
 + 2(\vec{x_{\alpha}}-\vec{x}_{\beta}) \cdot \left(
\frac{\vec{F}_{\alpha}}{m_{\alpha}}
-\frac{\vec{F}_{\beta}}{m_{\beta}}\right) \ ,
\end{equation}
where we can see that the calculation of $o^{I(\alpha,\beta)}$ takes always 
the same number of operations, independently of the number of atoms in our 
system, $n$, and the number of constraints imposed on it, $N_{c}$. Therefore, 
calculating the whole vector $o$ in eq.~(\ref{lm}) scales like $N_{c}$. 

In order to obtain an explicit expression for the entries of the matrix
$R$, we now introduce eq.~(\ref{grad_sigma}) into its definition in
eq.~(\ref{defR}):
\begin{eqnarray}
\label{sparseR}
R_{I(\alpha,\beta),J(\gamma,\epsilon)} & := & 
 \sum_{\zeta=1}^{n}{\frac{1}{m_{\zeta}}
   \vec{\nabla}_{\zeta}\sigma^{I(\alpha,\beta)} \cdot 
   \vec{\nabla}_{\zeta}\sigma^{J(\gamma,\epsilon)}} \nonumber \\
 & = & \sum_{\zeta=1}^{n}\frac{4}{m_{\zeta}}
      (\vec{x}_{\alpha}-\vec{x}_{\beta}) \cdot 
      (\vec{x}_{\gamma}-\vec{x}_{\epsilon})
      (\delta_{\zeta,\alpha}-\delta_{\zeta,\beta})
      (\delta_{\zeta,\gamma}-\delta_{\zeta,\epsilon}) \nonumber \\
 & = & 4 (\vec{x}_{\alpha}-\vec{x}_{\beta}) \cdot 
         (\vec{x}_{\gamma}-\vec{x}_{\epsilon})
  \left( \frac{\delta_{\alpha,\gamma}}{m_{\alpha}}
        -\frac{\delta_{\alpha,\epsilon}}{m_{\alpha}}
        -\frac{\delta_{\beta,\gamma}}{m_{\beta}}+
        \frac{\delta_{\beta,\epsilon}}{m_{\beta}}\right) \ ,
\end{eqnarray}
where we have used that
\begin{equation}
\sum_{\zeta=1}^{n}\delta_{\zeta,\alpha}\delta_{\zeta,\beta}
 =\delta_{\alpha,\beta} \ .
\end{equation}

Looking at this expression, we can see that a constant number of operations
(independent of $n$ and $N_{c}$) is required to obtain the value of every
entry in $R$. The terms proportional to the Kroenecker deltas imply that, as
we will see later, in a typical biological polymer, the matrix $R$ will be
sparse (actually banded if the constraints are appropriately ordered as we
describe in the following sections), being the number of non-zero entries
actually proportional to $N_c$. More precisely, the entry $R_{IJ}$ will only
be non-zero if the constraints $I$ and $J$ share an atom.

Now, since both the vector $o$ and the matrix $R$ in eq.~(\ref{lm}) can be
computed in $O(N_c)$ operations, it only remains to be proved that the
solution of the linear system of equations is also an $O(N_c)$ process, but
this is a well-known fact when the matrix defining the system is banded.
In~\cite{GR2010JCoP}, we introduced a new algorithm to solve this kind of
banded systems which is faster and more accurate than existing alternatives.
Essentially, we shown that the linear system of equations
\begin{equation}
\label{basic_system}
Ax=b \ ,
\end{equation}
where $A$ is a $d \times d$ matrix, $x$ is the $d \times 1$ vector of the 
unknowns, $b$ is a given $d \times 1$ vector and $A$ is \emph{banded}, i.e., 
it satisfies that for known $m<n$
\begin{eqnarray}
A_{I, I+K} &=& 0 \quad \forall \ K>m \ , \forall I  \ , \label{Aup}\\
A_{I+L,I} &=& 0 \quad \forall \ L>m \ , \forall I  \ , \label{Adown}
\end{eqnarray}
can be directly solved up to machine precision in $\mathcal{O}(d)$ operations.

This can be done using the following set of recursive equations for the
auxiliary quantities $\xi_{IJ}$ (see \cite{GR2010JCoP} for details):
\begin{subequations}
\label{coefsBand}
\begin{align}
\xi_{II} & = \left(A_{II} -
 \sum_{M=max(1,I-m)}^{I-1}\xi_{IM}\xi_{MI} \right)^{-1}
 \ , \label{xifinal_a} \\
\xi_{IJ} & = \xi_{II}\left(-A_{IJ} + 
 \sum_{M=\max\{1,J-m\}}^{I-1}\xi_{IM}\xi_{MJ}\right) \ ,
 & \textrm{for} \quad I < J \ , \label{xifinal_b} \\
\xi_{IJ} & = -A_{IJ} +
 \sum_{M=\max\{1,I-m\}}^{J-1}\xi_{IM}\xi_{MJ} \ ,
 & \textrm{for} \quad I > J \ , \label{xifinal_c}\\
 c_{I} & = b_{I} + \sum_{M=\max\{I-m,1\}}^{I-1}\xi_{IM}c_{M} \ , \label{ci} \\
 x_{I} & = \xi_{II}c_{I}+\sum_{K=I+1}^{\min\{I+m,n\}}\xi_{IK}x_{K}
   \label{xi} \ .
\end{align}
\end{subequations}

If the matrix $A$ is symmetric ($A_{IJ}=A_{JI}$), as it is the case with $R$
[see~(\ref{defR})], we can additionally save about one half of the computation
time just by using
\begin{equation}
\xi_{IJ}=\xi_{JI}/\xi_{JJ} \ , \qquad \textrm{for} \quad I > J
 \ , \label{xifinal_csim} 
\end{equation}
instead of (\ref{xifinal_c}). Eq. (\ref{xifinal_csim}) can be obtained from
(\ref{coefsBand}) by induction, and we recommend these expressions for the
$\xi$ coefficients because other valid ones (like considering
$\xi_{IJ}=\xi_{JI}$, $\xi_{II}=1/\sqrt{A_{II} -
\sum_{M=max(1,I-m)}^{I-1}\xi_{IM}\xi_{MI}}$, which involves square roots) are
computationally more expensive.

In the next sections, we show how to index the constraints in such a way that
nearby indices correspond to constraints where involved atoms are close to
each other and likely participate of the same constraints. In such a case, not
only will the matrix $R$ in eq.~(\ref{lm}) be banded, allowing to use the
method described above, but it will also have a minimal bandwidth $m$, which
is also an important point, since the computational cost for solving the
linear system scales as $\mathcal{O}(N_{c}m^{2})$ (when the bandwidth is
constant).

\section{Ordering of the constraints}
\label{soc}

In this section we describe how to index the constraints applied to the
bond lengths and bond angles of a series of model systems and biological
molecules with the already mentioned aim of minimizing the computational
cost associated to the obtention of the Lagrange multipliers. The
presentation begins by deliberately simple systems and proceeds to
increasingly more complicated molecules with the intention that the reader
is not only able to use the final results presented here, but also
to devise appropriate indexings for different molecules not covered in this
work.

The main idea we have to take into account, as expressed in section
\ref{sec_aclm}, is to use nearby numbers to index constraints containing the
same atoms. If we do so, we will obtain \emph{banded} $R$ matrices. Further
computational savings can be obtained if we are able to reduce the number of
$\xi$ coefficients in eqs.~(\ref{coefsBand}) to be calculated. In more detail,
solving a linear system like~(\ref{lm}) where the $R$ is $N_c \times N_c $ and
banded with semi-band width (i.e., the number of non-zero entries neighbouring
the diagonal in one row or column) $m$ requires $O(N_c m^2)$ operations if $m$
is a constant. Therefore, the lower the value of $m$, the smaller the number
of required numerical effort. When the semi-band width $m$ is not constant
along the whole matrix, things are more complicated and the cost is always
between $O(N_c m_\mathrm{min}^2)$ and $O(N_c m_\mathrm{max}^2)$, depending on
how the different rows are arranged. In general, we want to minimize the
number of zero fillings in the process of Gaussian elimination (see
\cite{GR2010JCoP} for further details), which is achieved by not having zeros
below non-zero entries.

This is easier to understand with an example: Consider the following matrices,
where $\Omega$ and $\omega$ represent different non-zero values for every
entry (i.e., not all $\omega$, nor all $\Omega$ must take the same value,
and different symbols have been chosen only to highlight the main diagonal):
\begin{equation}
A := \left(
\begin{array}{cccccc}
 \Omega & \omega & \omega & \omega & 0 & \ldots\\
       & \Omega & 0      &      0 & 0 & \ldots
\end{array}
     \right) \ , \quad
B :=
  \left(
\begin{array}{cccccc}
 \Omega & \omega & \omega & 0      & 0 & \ldots \\
       & \Omega & \omega &      0 & 0 & \ldots 
\end{array}
  \right) \ .
\end{equation}

During the Gaussian elimination process that is behind (\ref{coefsBand}), in
$A$, five coefficients $\xi$ above the diagonal are to be calculated, three in
the first row and two in the second one, because the entries below non-zero
entries become non-zero too as the elimination process advances (this is what
we have called `zero filling'). On the other hand, in $B$, which contains the
same number of non-zero entries as $A$, only three coefficients $\xi$ have to
be calculated: two in the first row and one in the second row. Whether $R$
looks like $A$ or like $B$ depends on our choice of the constraints ordering.

One has also to take into account that no increase in the computational cost
occurs if a series of non-zero columns is separated from the diagonal by
columns containing all zeros. I.e., the linear systems associated to the
following two matrices require the same numerical effort to be solved:
\begin{equation}
C := \left(
\begin{array}{ccccccccccc}
 \Omega & \omega & \omega & 0 & 0  & \omega & \omega & 0 &  \ldots\\
       & \Omega & \omega  & 0 & 0  &\omega & \omega & 0 &  \ldots \\
       &       &  \Omega  & 0 & 0  &\omega & \omega & 0 &  \ldots 
\end{array}
   \right) \ , \quad 
D := \left(
\begin{array}{cccccccc}
 \Omega & \omega & \omega   & \omega & \omega & 0 &  \ldots\\
       & \Omega & \omega   &\omega & \omega & 0 &  \ldots \\
       &       &  \Omega    &\omega & \omega & 0 &  \ldots 
\end{array}
   \right) \ .
\end{equation}

\subsection{Open, single-branch chain with constrained bond lengths}
\label{secBL}

\begin{figure}[b!]
\centering
\includegraphics[width=9cm]{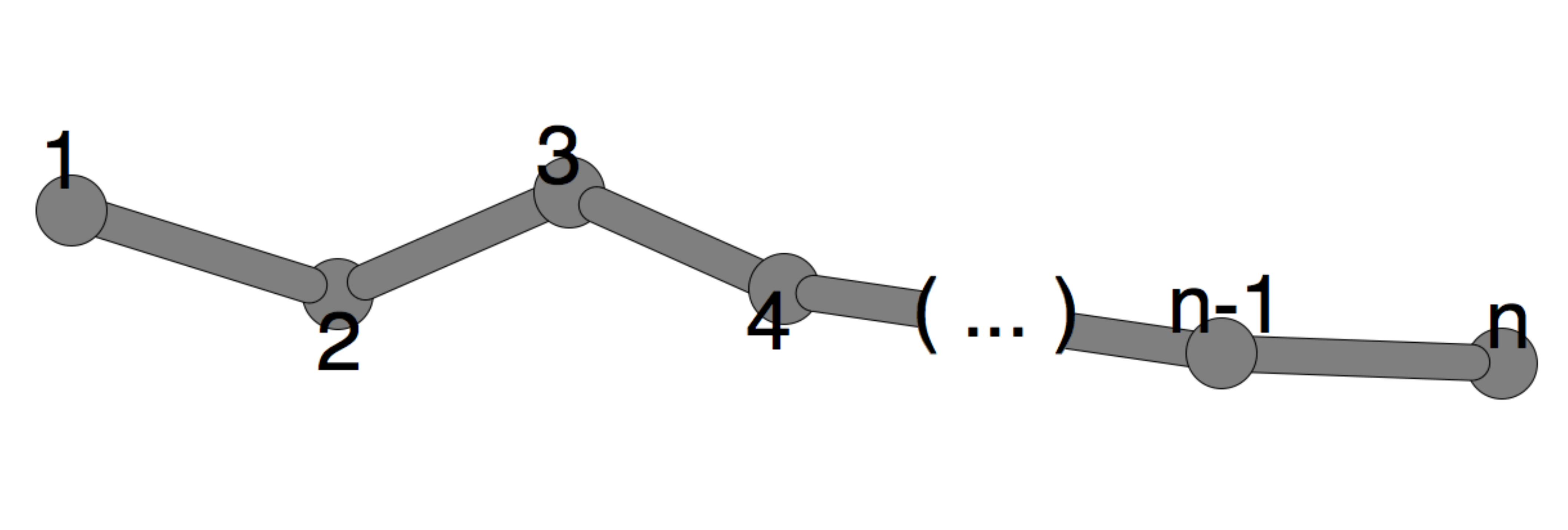}
\caption{Numbering of the atoms in an open, single-branch chain.}
\label{fig:lc}
\end{figure}

As promised, we start by a simple model of a biomolecule: an open linear chain
without any branch. In this case, the atoms should be trivially numbered
as in fig.~\ref{fig:lc} (any other arrangement would have to be justified
indeed!).

If we only constrain bond lengths, the fact that only consecutive atoms
participate of the same constraints allows us to simplify the notation with
respect to eq.~(\ref{sigma_generica}) and establish the following ordering
for the constraints indices:
\begin{equation}
\label{blc1}
I(\alpha)=\alpha \ , \quad I=1,\ldots,n-1 \ ,
\end{equation}
with
\begin{equation}
\label{blc1_constr}
\sigma^{I(\alpha)}(\vec{x}_{\alpha},\vec{x}_{\alpha+1})
 :=(\vec{x}_{\alpha}-\vec{x}_{\alpha+1})^{2}-(a_{\alpha,\alpha+1})^{2}=0  
 \ .
\end{equation}

This choice results in a tridiagonal matrix $R$, whose only non-zero entries
are those lying in the diagonal and its first neighbours. This is the only
case for which an exact calculation of the Lagrange multipliers exists in the
literature as far as we are aware \cite{Maz2007JPAMT}.

\subsection{Open, single-branch chain with constrained bond lengths and bond 
            angles}
\label{angles}

The next step in complexity is to constrain the bond angles of the same linear
chain that we discussed above. The atoms are ordered in the same way, as in
fig. \ref{fig:lc}, and the trick to generate a banded matrix $R$ with minimal
bandwidth is to alternatively index bond length constraints with odd numbers,
\begin{equation}
I(\alpha)=2\alpha-1=1,3,5,\ldots,2n-3 \ , \quad \mathrm{with} \quad
  \alpha=1,2,\ldots,n-1 \ ,
\end{equation}
and bond angle constraints with even ones,
\begin{equation}
J(\beta)=2\beta=2,4,6,\ldots,2n-4 \ , \quad \mathrm{with} \quad
  \beta=1,2,\ldots,n-2 \ ,
\end{equation}
where the regular pattern involving the atom indicies that participate of
the same constraints has allowed again to use a lighter notation.

The constraints equations in this case are
\begin{subequations}
\begin{align}
& \sigma^{I(\alpha)}(\vec{x}_{\alpha},\vec{x}_{\alpha+1})
 =(\vec{x}_{\alpha}-\vec{x}_{\alpha+1})^{2}-(a_{\alpha,\alpha+1})^{2}=0 \ ,
 \\
& \sigma^{J(\beta)}(\vec{x}_{\beta},\vec{x}_{\beta+2})
  =(\vec{x}_{\beta}-\vec{x}_{\beta+2})^{2}-(a_{\beta,\beta+2})^{2}=0 \ ,
\end{align}
\end{subequations}
respectively, and, if this indexing is used, $R$ is a banded matrix where $m$
is 3 and 4 in consecutive rows and columns. Therefore, the mean $\langle m 
\rangle$ is 3.5, and the number of $\xi$ coefficients that have to be
computed per row in the Gaussian elimination process is the same because
the matrix contains no zeros that are filled.

\begin{figure}[!t]
\centering
\includegraphics[width=7cm]{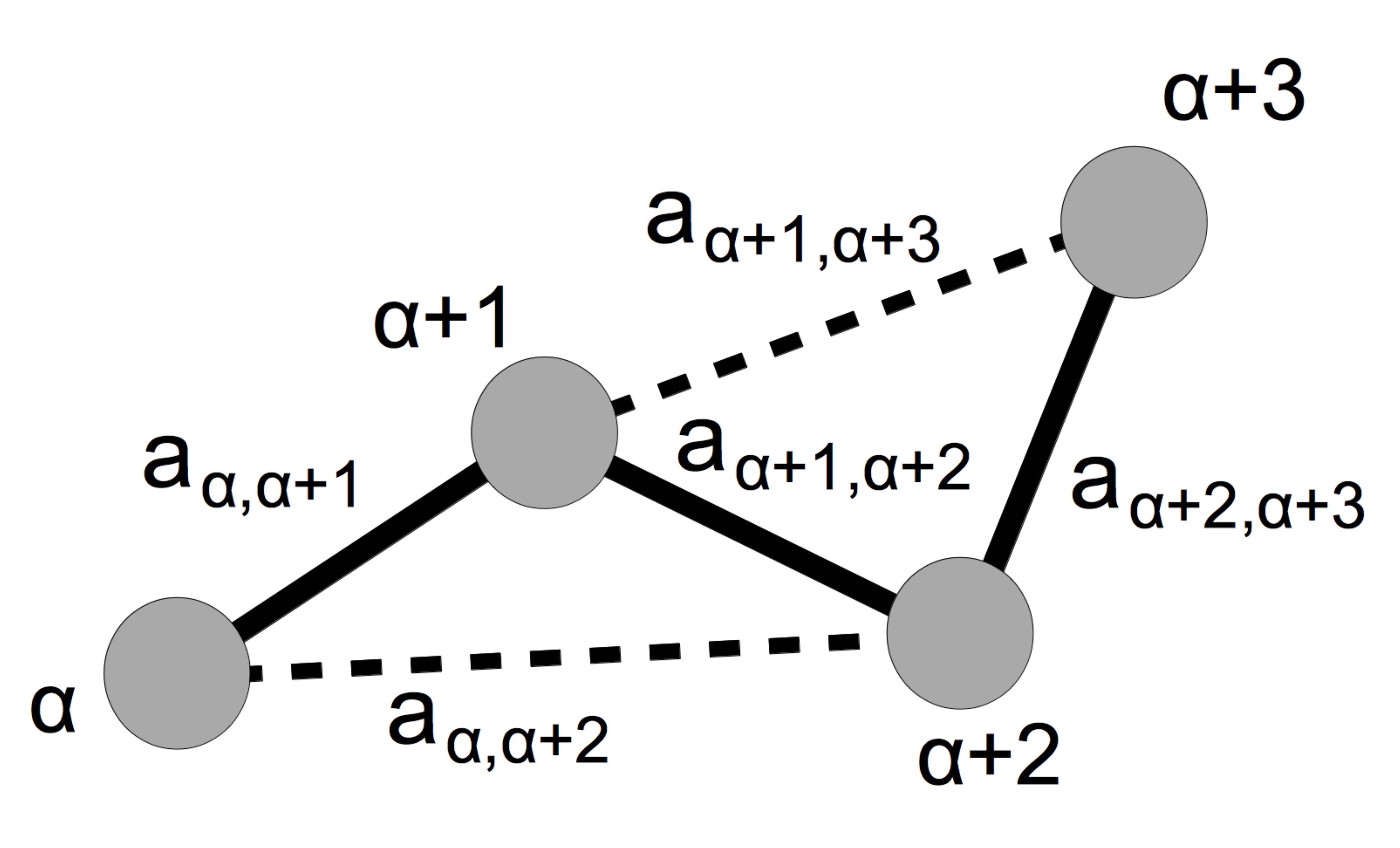}
\caption{Segment of a single-branch chain with constrained bond lengths and
bond angles. In solid line, the distances that have to be kept constant to
constrain the former; in dashed line, those that have to be kept constant to
constrain the latter.}
\label{fig:angs}
\end{figure}

A further feature of this system (and other systems where both bond lengths
and bond angles are constrained) can be taken into account in order to reduce
the computational cost of calculating Lagrange multipliers in a molecular
dynamics simulation: A segment of the linear chain with constrained bond
lengths and bond angles is represented in fig.~\ref{fig:angs}, where the
dashed lines correspond to the virtual bonds between atoms that, when kept
constant, implement the constraints on bond angles (assuming that the bond
lengths, depicted as solid lines, are also constrained).

Due to the fact that all these distances are constant, many of the entries of
$R$ will remain unchanged during the molecular dynamics simulation. As an
example, we can calculate
\begin{eqnarray}
\label{ejang}
R_{2\alpha-1,2\alpha-2} & = &
\frac{1}{m_{\alpha}}(\vec{x}_{\alpha}-\vec{x}_{\alpha+1})\cdot
   (\vec{x}_{\alpha}-\vec{x}_{\alpha+2}) =
\frac{a_{\alpha,\alpha+1}a_{\alpha,\alpha+2}}
     {m_{\alpha}}\cos\angle{(\alpha+1,\alpha,\alpha+2)} \nonumber \\
 & = & \frac{1}{2m_{\alpha}}
  (a_{\alpha,\alpha+1}^{2}+(a_{\alpha,\alpha+2})^{2}-
   a_{\alpha+1,\alpha+2}^{2}) \ .
\end{eqnarray}
where we have used the law of cosines. The right-hand side does not depend on
any time-varying objects (such as $\vec{x}_{\alpha}$), being made of only
constant quantities. Therefore, the value of $R_{2\alpha-1,2\alpha-2}$ (and
many other entries) needs not to be recalculated in every time step, which
allows to save computation time in a molecular dynamics simulation.

\subsection{Minimally branched molecules with constrained bond lengths}
\label{sec:singly_branched}

\begin{figure}[!t]
\centering
\includegraphics[width=11cm]{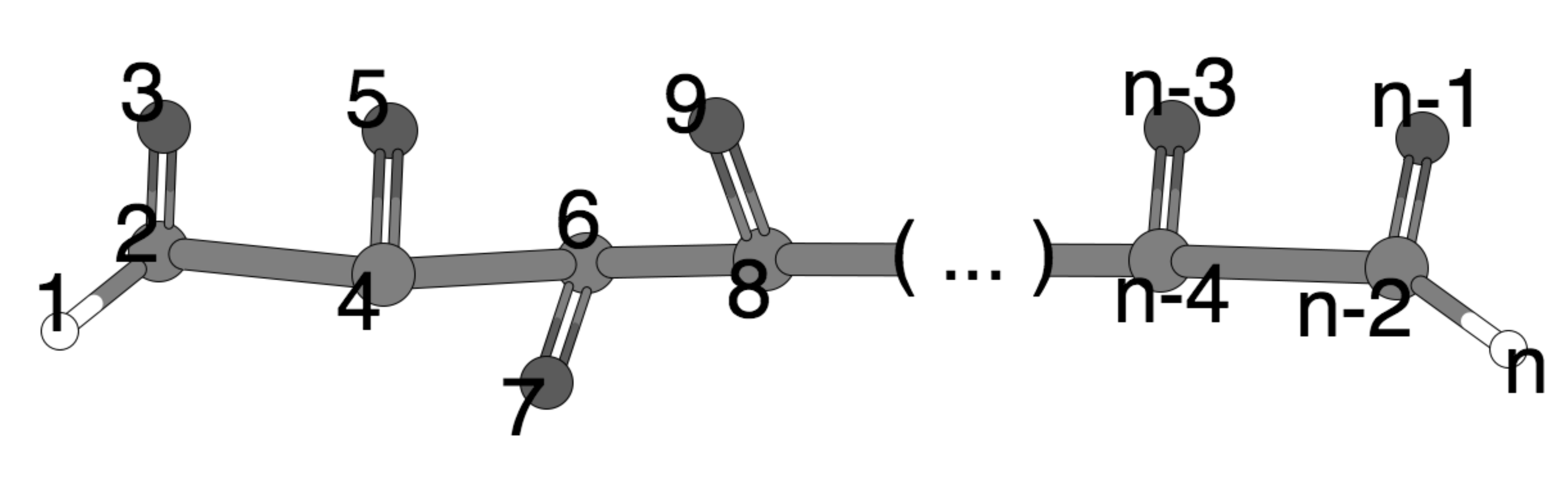}
\caption{Numbering of the atoms in a minimally branched molecule.}
\label{fig:branched1}
\end{figure}

In order to incrementally complicate the calculations, we now turn to a
linear molecule with only one atom connected to the backbone, such the one 
displayed in figure \ref{fig:branched1}.

The corresponding equations of constraint and the ordering in the indices
that minimizes the bandwidth of the linear system are
\begin{subequations}
\label{singlybranched}
\begin{align}
&\sigma^{1} = (\vec{x}_{1}-\vec{x}_{2})^{2}-a_{1,2}^{2}=0  \ , \\
&\sigma^{I(\alpha)} = (\vec{x}_{\alpha}-\vec{x}_{\alpha+1})^{2}
 -a_{\alpha,\alpha+1}^{2}=0 \ , \quad
 I(\alpha)=\alpha=2,4,6,\ldots,n-2 \ , \\
&\sigma^{J(\beta)} = (\vec{x}_{\beta}-\vec{x}_{\beta+2})^{2}
 -a_{\beta,\beta+2}^{2}=0  \ , \quad
 J(\beta)=\beta+1=3,5,7,\ldots,n-1 \ ,
\end{align}
\end{subequations}
where the trick this time has been to alternatively consider atoms in the
backbone and atoms in the branches as we proceed along the chain.

The matrix $R$ of this molecule presents a semi-band width which is
alternatively 2 and 1 in consecutive rows/columns, with average $\langle m
\rangle=1.5$ and the same number of superdiagonal $\xi$ coefficients to be
computed per row.

\subsection{Alkanes with constrained bond lengths}
\label{sec:doubly_branched}

\begin{figure}[b!]
\centering
\includegraphics[width=11cm]{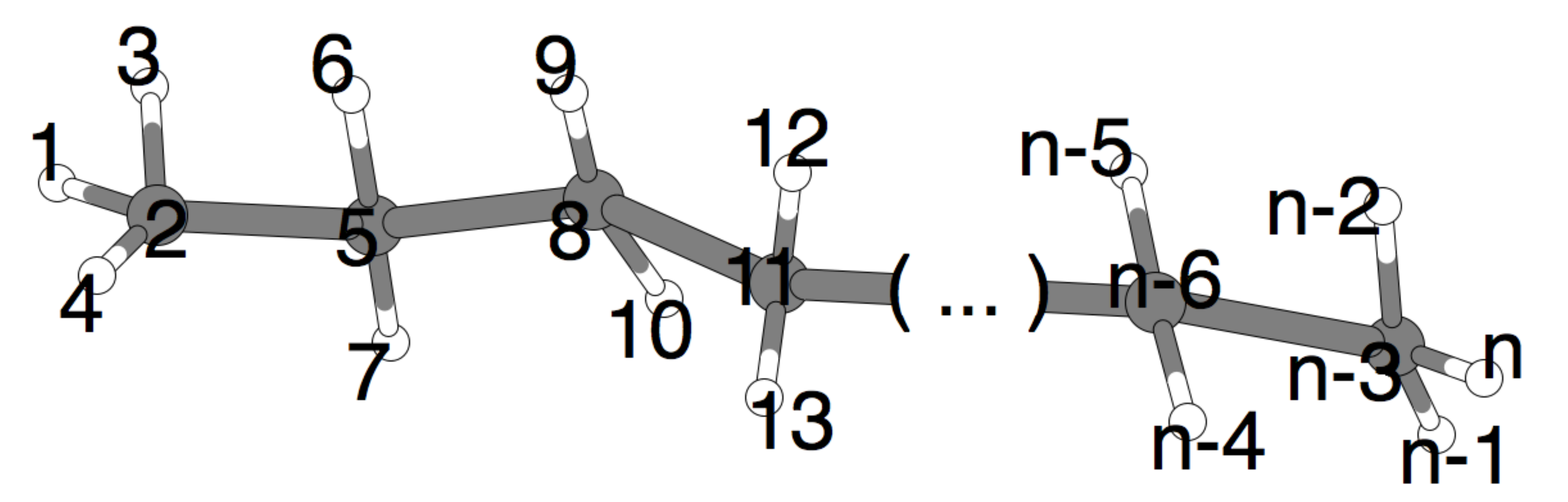}
\caption{Numbering of the atoms in a model alkane chain.}
\label{fig:nalkane}
\end{figure}

The next molecular topology we will consider is that of an alkane
(a familiy of molecules with a long tradition in the field of constraints
\cite{Ryc1977JCOP}), i.e., a linear backbone with two 1-atom branches
attached to each site (see fig.~\ref{fig:nalkane}).

The ordering of the constraints that minimizes the bandwidth of the linear
system for this case is
\begin{subequations}
\begin{align}
& \sigma^{1}(x) = (\vec{x}_{1}-\vec{x}_{2})^{2}-a_{1,2}^{2}=0 \ , \\
& \sigma^{I(\alpha)}(x) = (\vec{x}_{\alpha}-\vec{x}_{\alpha+3})^{2}
  -a_{\alpha+3,\alpha}^{2}=0 \ , \quad
  I(\alpha)=\alpha=2,5,8,\ldots,n-3 \ , \\
& \sigma^{J(\beta)}(x) = (\vec{x}_{\beta}-\vec{x}_{\beta+1})^{2}
  -a_{\beta+1,\beta}^{2}=0 \ , \quad
  J(\beta)=\beta+1=3,6,9,\ldots,n-2 \ , \\
& \sigma^{K(\gamma)}(x) = (\vec{x}_{\gamma}-\vec{x}_{\gamma+2})^{2}
  -a_{\gamma+2,\gamma}^{2}=0 \ , \quad
  K(\gamma)=\gamma+2=4,7,10,\ldots,n-1 \ ,
\end{align}
\end{subequations}
where the trick has been in this case to alternatively constrain the bond
lengths in the backbone and those connecting the branching atoms to one side
or the other. The resulting $R$ matrix require the calculation of 2
$\xi$ coefficients per row when solving the linear system.

\subsection{Minimally branched molecules with constrained bond lengths and 
            bond angles}
\label{sbblba}

If we want to additionally constrain bond angles in a molecule with the 
topology in fig.~\ref{fig:branched1}, the following ordering is convenient:
\begin{subequations}
\begin{align}
& \sigma^{1}(x)=(\vec{x}_{1}-\vec{x}_{2})^{2}-a_{1,2}^{2}=0 \ , \\
& \sigma^{2}(x)=(\vec{x}_{2}-\vec{x}_{3})^{2}-a_{2,3}^{2}=0 \ , \\
& \sigma^{3}(x)=(\vec{x}_{1}-\vec{x}_{4})^{2}-a_{1,4}^{2}=0 \ , \\
& \sigma^{I(\alpha)}(x)=(\vec{x}_\alpha-\vec{x}_{\alpha+1})^{2}-
  a_{\alpha,\alpha+1}^{2}=0 \ , \quad
  I(\alpha)=2\alpha-2=4,8,12,\ldots,2n-4 \ , \\
& \sigma^{J(\beta)}(x)=(\vec{x}_\beta-\vec{x}_{\beta+2})^{2}-
  a_{\beta,\beta+2}^{2}=0 \ , \quad
  J(\beta)=2\beta+2=5,9,13,\ldots,2n-3 \ , \\
& \sigma^{K(\gamma)}(x)=(\vec{x}_\gamma-\vec{x}_{\gamma+1})^{2}-
  a_{\gamma,\gamma+1}^{2}=0 \ , \quad
  K(\gamma)=2\gamma-2=6,10,14,\ldots,2n-6 \ , \\
& \sigma^{L(\delta)}(x)=(\vec{x}_\delta-\vec{x}_{\delta+4})^{2}-
  a_{\delta,\delta+4}^{2}=0 \ , \quad
  L(\delta)=2\delta+3=7,11,15,\ldots,2n-5 \ .
\end{align}
\end{subequations}

This ordering produces 16 non-zero entries above the diagonal per each group
of 4 rows in the matrix $R$ when making the calculations to solve the
associated linear system. This is, we will have to calculate a mean of
$16/4=4$ super-diagonal coefficients $\xi$ per row. When we studied the linear
molecule with constrained bond lengths and bond angles, this mean was equal to
$3.5$, so including minimal branches in the linear chain makes the
calculations just slightly longer.

\subsection{Alkanes with constrained bond lengths and bond angles}
\label{nalk}

If we now want to add bond angle constraints to the bond length ones described
in sec.~\ref{sec:doubly_branched} for alkanes, the following ordering
produces a matrix $R$ with a low half-band width:
\begin{subequations}
\begin{align}
& \sigma^{1}(x)=(\vec{x}_{2}-\vec{x}_{1})^{2}-a_{1,2}^{2}=0 \ , \\
& \sigma^{2}(x)=(\vec{x}_{3}-\vec{x}_{2})^{2}-a_{2,3}^{2}=0 \ , \\
& \sigma^{3}(x)=(\vec{x}_{4}-\vec{x}_{2})^{2}-a_{2,4}^{2}=0 \ , \\
& \sigma^{4}(x)=(\vec{x}_{5}-\vec{x}_{1})^{2}-a_{1,5}^{2}=0 \ , \\
& \sigma^{5}(x)=(\vec{x}_{5}-\vec{x}_{3})^{2}-a_{3,5}^{2}=0 \ , \\
& \sigma^{6}(x)=(\vec{x}_{5}-\vec{x}_{4})^{2}-a_{4,5}^{2}=0 \ , \\
& \sigma^{I(\alpha)}(x)=(\vec{x}_\alpha-\vec{x}_{\alpha+3})^{2}-
  a_{\alpha,\alpha+3}^{2}=0 \ , \quad
  I(\alpha)=2\alpha+3=7,13,19,\ldots,2n-3 \ , \\
& \sigma^{J(\beta)}(x)=(\vec{x}_\beta-\vec{x}_{\beta+1})^{2}-
  a_{\beta,\beta+1}^{2}=0 \ , \quad
  J(\beta)=2\beta-2=8,14,20,\ldots,2n-8 \ , \\
& \sigma^{K(\gamma)}(x)=(\vec{x}_\gamma-\vec{x}_{\gamma+2})^{2}-
  a_{\gamma,\gamma+2}^{2}=0 \ , \quad
  K(\gamma)=2\gamma-1=9,15,21,\ldots,2n-7 \ , \\
& \sigma^{L(\delta)}(x)=(\vec{x}_\delta-\vec{x}_{\delta+6})^{2}-
  a_{\delta,\delta+6}^{2}=0 \ , \quad
  L(\delta)=2\delta+6=10,16,22,\ldots,2n-6 \ , \\
& \sigma^{M(\epsilon)}(x)=(\vec{x}_\epsilon-\vec{x}_{\epsilon+2})^{2}-
  a_{\epsilon,\epsilon+2}^{2}=0 \ , \quad
  M(\epsilon)=2\epsilon-1=11,17,23,\ldots,2n-5 \ ,  \\
& \sigma^{N(\zeta)}(x)=(\vec{x}_\zeta-\vec{x}_{\zeta+1})^{2}-
  a_{\zeta,\zeta+1}^{2}=0 \ , \quad
  N(\zeta)=2\zeta-2=12,18,24,\ldots,2n-4 \ .
\end{align}
\end{subequations}

In this case, the average number of $\xi$ coefficients to be calculated
per row is approximately 5.7.

\begin{figure}[t!]
\centering
\includegraphics[width=10cm]{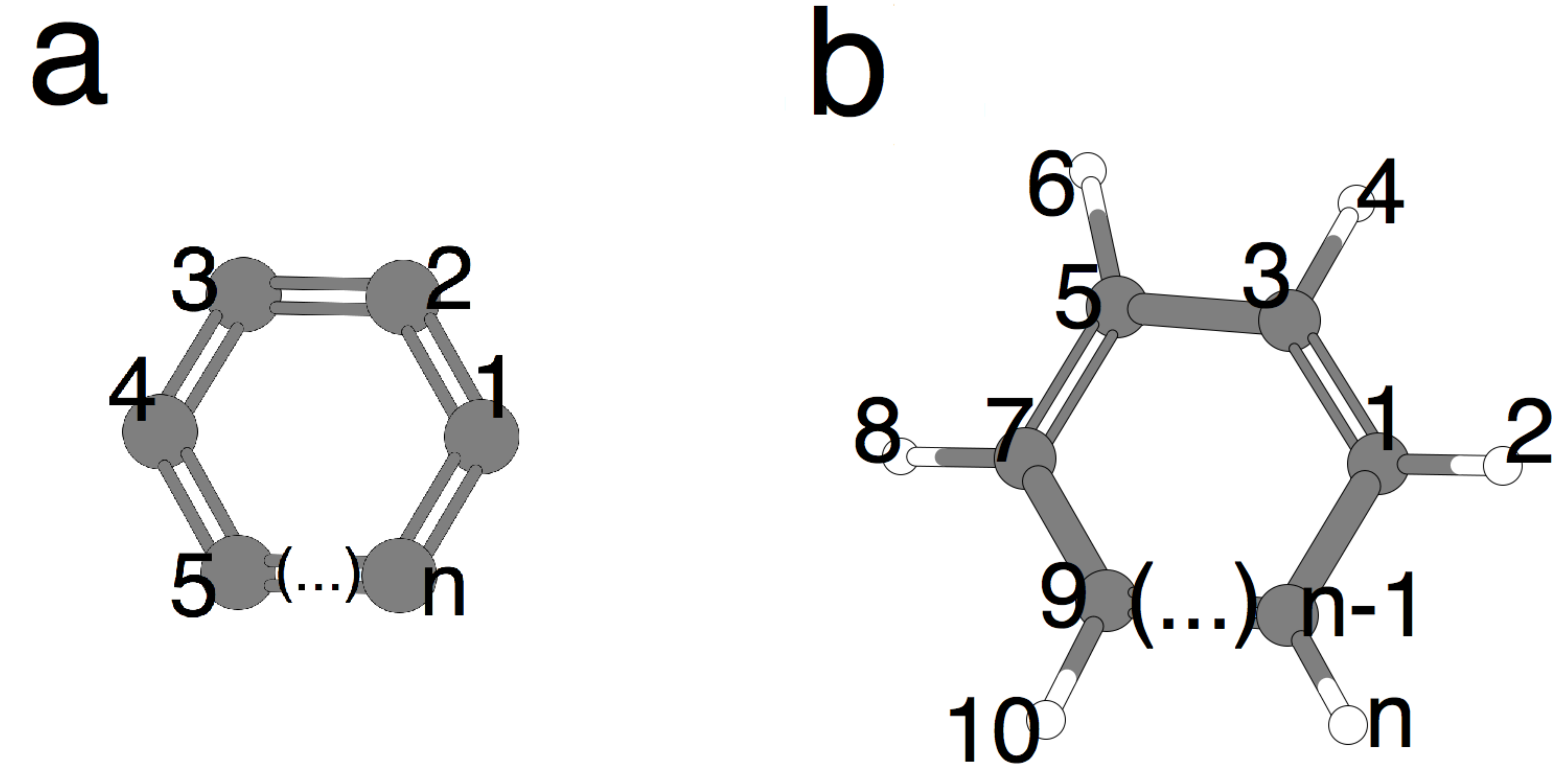}
\caption{Numbering of the atoms for cyclic molecules: \textbf{a)} without 
branches; \textbf{b)} minimally branched.}
\label{fig:rings}
\end{figure}

\subsection{Cyclic chains}
\label{ring1}

If we have cycles in our molecules, the indexing of the constraints is only
slightly modified with respect to the open cases in the previous sections. For
example, if we have a single-branch cyclic topology, such as the one displayed
in fig.~\ref{fig:rings}a, the ordering of the constraints is the following:
\begin{subequations}
\begin{align}
& \sigma^{I(\alpha)}(x)=(\vec{x}_{\alpha}-\vec{x}_{ \alpha+1})^{2}
   -(a_{\alpha,\alpha+1})^{2}=0 \ ,
  \quad I(\alpha)=1,\ldots,n-1=\alpha \ , \\
& \sigma^{n}(x) = (\vec{x}_{1}-\vec{x}_{n})^{2}-(a_{1,n})^{2} = 0
  \ . \label{constrcerrar}
\end{align}
\end{subequations}

These equations are the same as those in \ref{secBL}, plus a final constraint
corresponding to the bond which closes the ring. These constraints produce a
matrix $R$ where only the diagonal entries, its first neighbours, and the
entries in the corners ($R_{1,n}$ and $R_{n,1}$) are non-zero. In this case,
the associated linear system in eq.~(\ref{lm}) can also be solved in $O(N_c)$
operations, as we discuss in \cite{GR2010JCoP}. In general, this is also valid
whenever $R$ is a sparse matrix with only a few non-zero entries outside of
its band, and therefore we can apply the technique introduced in this work to
molecular topologies containing more than one cycle.

The ordering of the constraints and the resulting linear systems for different
cyclic species, such as the one depicted in fig.~\ref{fig:rings}b, can be
easily constructed by the reader using the same basic ideas.

\subsection{Proteins}
\label{sec:proteins}

As we discussed in sec.~\ref{sec:introduction}, proteins are one of the most
important families of molecules from the biological point of view: Proteins
are the nanomachines that perform most of the complex tasks that need to be
done in living organisms, and therefore it is not surprising that they are
involved, in one way or another, in most of the diseases that affect animals
and human beings. Given the efficiency and precision with which proteins carry
out their missions, they are also being explored from the technological point
of view. The applications of proteins even outside the biological realm are
many if we could harness their power \cite{Ech2007COP}, and molecular dynamics
simulations of great complexity and scale are being done in many laboratories
around the world as a tool to understand them
\cite{Luc2007PNAS,Sha2009XXX,Fre2009BJ}.

Proteins present two topological features that simplify the calculation of the
Lagrange multipliers associated to constraints imposed on their degrees of
freedom:

\begin{figure}[!t]
\centering
\includegraphics[width=8cm]{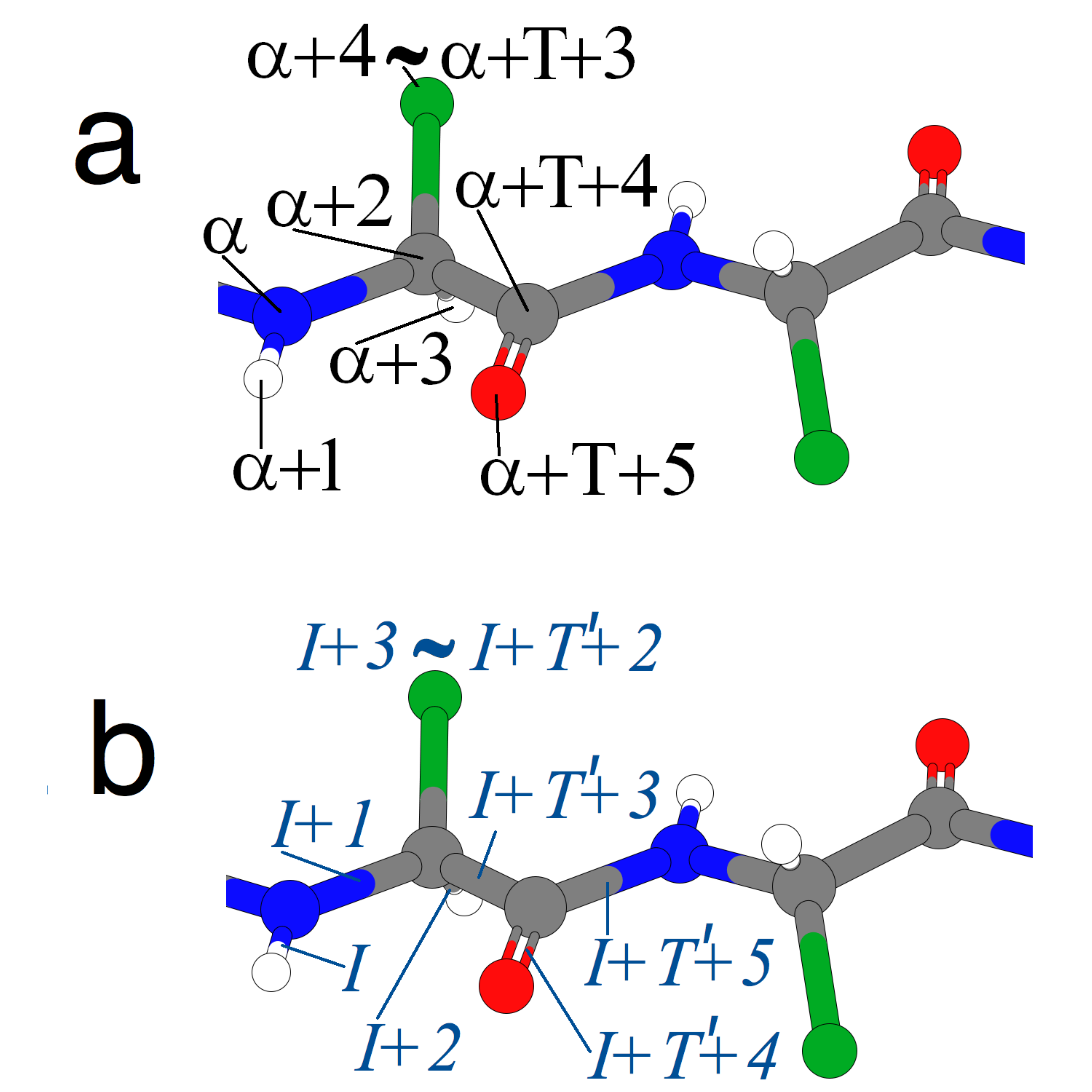}
\caption{Scheme of the residue of a protein. \textbf{a)} Numbering of the
atoms; $\alpha$ represents the first numbered atom in each residue (the amino
nitrogen) and $T$ is the number of atoms in the side chain. \textbf{b)}
Indexing of the bond length constraints; $I$ denotes the index of the first
constraint imposed on the residue (the N-H bond length) and $T^\prime$ is the
variable number of constraints imposed on the side chain.}
\label{fig:prot}
\end{figure}

\begin{itemize}

\item They are linear polymers, consisting of a backbone with short (17 atoms 
at most) groups attached to it \cite{Ech2007COP}. This produces a banded matrix $R$, thus allowing the solution of the associated linear problem in $O(N_c)$ operations. Even in the case that disulfide bridges, or any other
covalent linkage that disrupts the linear topology of the molecule, exist,
the solution of the problem can still be found efficiently if we recall
the ideas discussed in sec.~\ref{ring1}.

\item The monomers that typically make up these biological polymers, i.e., the
residues associated to the proteinogenic aminoacids, are only 20 different
molecular structures. Therefore, it is convenient to write down explicitly one
block of the $R$ matrix for each known monomer, and to build the $R$ matrix of
any protein simply joining together the precalculated blocks associated to the
corresponding residues the protein consists of.

\end{itemize}

The structure of a segment of the backbone of a protein chain is depicted in
fig.~\ref{fig:prot}. The green spheres represent the side chains, which are
the part of the amino acid residue that can differ from one monomer to the
next, and which usually consist of several atoms: from 1 atom in the case of
glycine to 17 in arginine or tryptophan. In fig.~\ref{fig:prot}a, we present
the numbering of the atoms, which will support the ordering of the
constraints, and, in fig.~\ref{fig:prot}b, the indexing of the constraints is
presented for the case in which only bond lengths are constrained (the bond
lengths plus bond angles case is left as an exercise for the reader).

Using the same ideas and notation as in the previous sections and denoting by
$R_\mathcal{M}$ the block of the matrix $R$ that corresponds to a given amino
acid residue $\mathcal{M}$, with
$\mathcal{M}=1,\ldots,\mathcal{N}_\mathcal{R}$, we have that, for the monomer
dettached of the rest of the chain,
\begin{equation}
\label{Rpart}
R_\mathcal{M} = 
\left(
\begin{array}{ccc|lcr|ccc}
 \Omega & \omega & & & & & & &  
\\
\omega &\Omega & \omega & \omega & & & \omega & &  
\\
 & \omega  & \Omega & \omega & & & \omega & &   
\\
\hline
 &\omega & \omega & & & &  \omega &  & \\ 
 & & & & S & & & & \\ 
 & & & & & & & & 
\\ \hline
 &\omega& \omega &\omega & &  & \Omega & \omega & \omega  \\
 & & & & &  & \omega & \Omega & \omega \\
 & & & & &  & \omega &\omega  & \Omega \\
\end{array} 
\right) \ ,
\end{equation}
where the explicit non-zero entries are related to the constraints imposed on
the backbone and $S$ denotes a block associated to those imposed on the bonds
that belong to the different sidechains. The dimension of this matrix is
$T^\prime+6$ and the maximum possible semi-band width is 12 for the bulkiest
residues.

\begin{figure}[!hb]
\centering
\includegraphics[width=7cm]{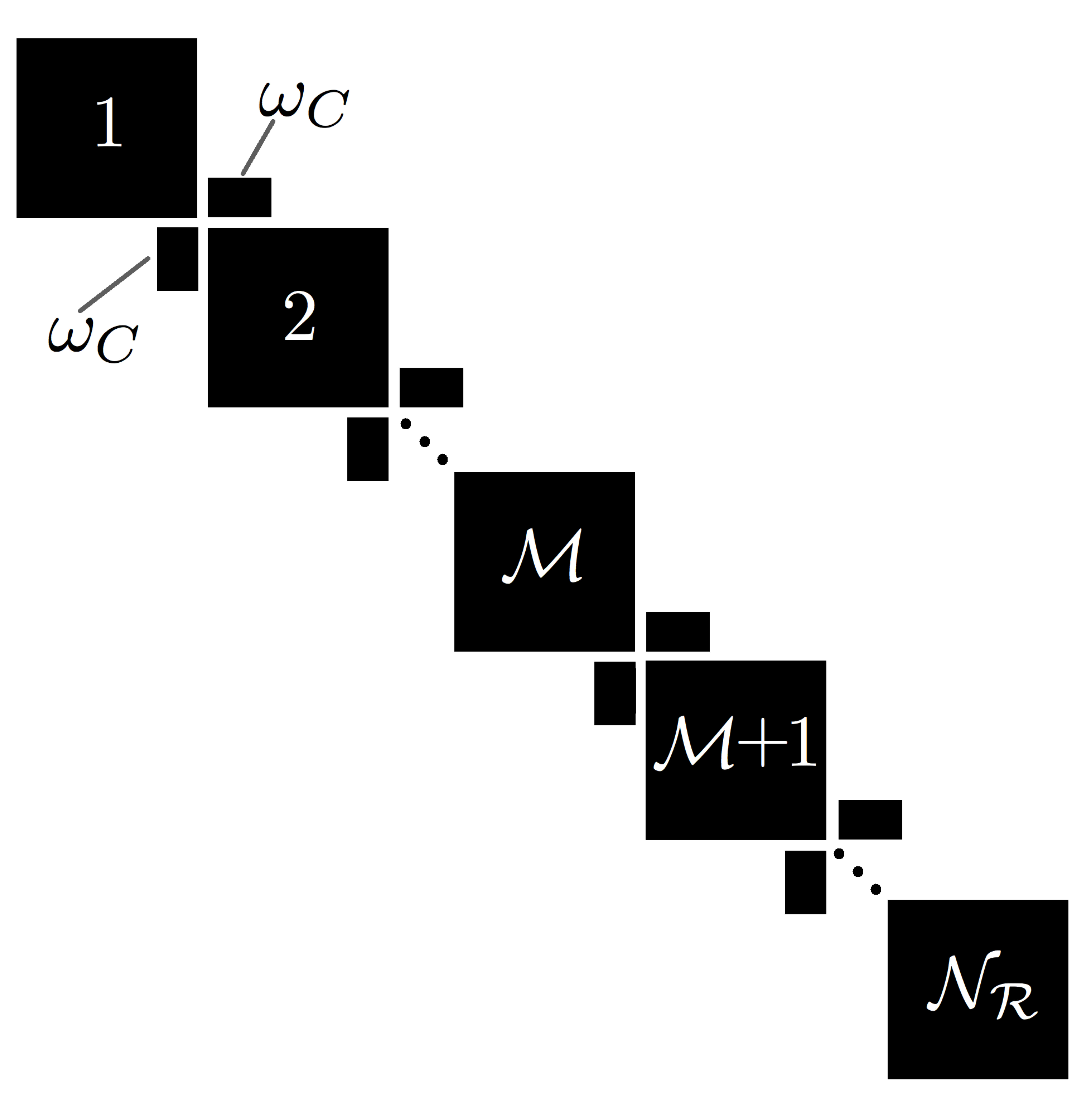}
\caption{Scheme of the matrix $R$ for a protein molecule with $\mathcal{N_R}$ 
residues. In black, we represent the potentially non-zero entries, and each 
large block in the diagonal is given by~(\ref{defR}).}
\label{fig:proteinmatrix}
\end{figure}

A protein's global matrix $R$ has to be built by joining together blocks like
the one above, and adding the non-zero elements related to the imposition of
constraints on bond lengths that connect one residue with the next. These
extra elements are denoted by $\omega_C$ and a general scheme of the final
matrix is shown in fig.~\ref{fig:proteinmatrix}.

The white regions in this scheme correspond to zero entries, and we can easily
check that the matrix is banded. In fact, if each one of the diagonal blocks
is constructed conveniently, they will contain many zeros themselves and the
bandwidth can be reduced further. The size of the $\omega_{C}$ blocks will
usually be much smaller than that of their neighbour diagonal blocks. For
example, in the discussed case in which we constrain all bond lengths,
$\omega_{C}$ are $1 \times 2$ (or $2 \times 1$) blocks, and the diagonal
blocks size is between $7 \times 7$ (glycine) and $25 \times 25$ (tryptophan).

\subsection{Nucleic acids}
\label{sec:nucleic_acids}

Nucleic acids are another family of very important biological molecules that
can be tackled with the techniques described in this work. DNA and RNA, the
two subfamilies of nucleic acids, consist of linear chains made up of a finite
set of monomers (called `bases'). This means that they share with proteins the
two features mentioned in the previous section and therefore the Lagrange
multipliers associated to the imposition of constraints on their degrees of
freedom can be efficiently computed using the same ideas. It is worth
mentioning that DNA typically appears in the form of two complementary chains
whose bases form hydrogen-bonds. Since these bonds are much weaker than a
covalent bond, imposing bond length constraints on them such as the ones in
eq.~(\ref{sigma_generica}) would be too unrealistic for many practical
purposes,
 
\begin{figure}[b!]
\centering
\includegraphics[width=9cm]{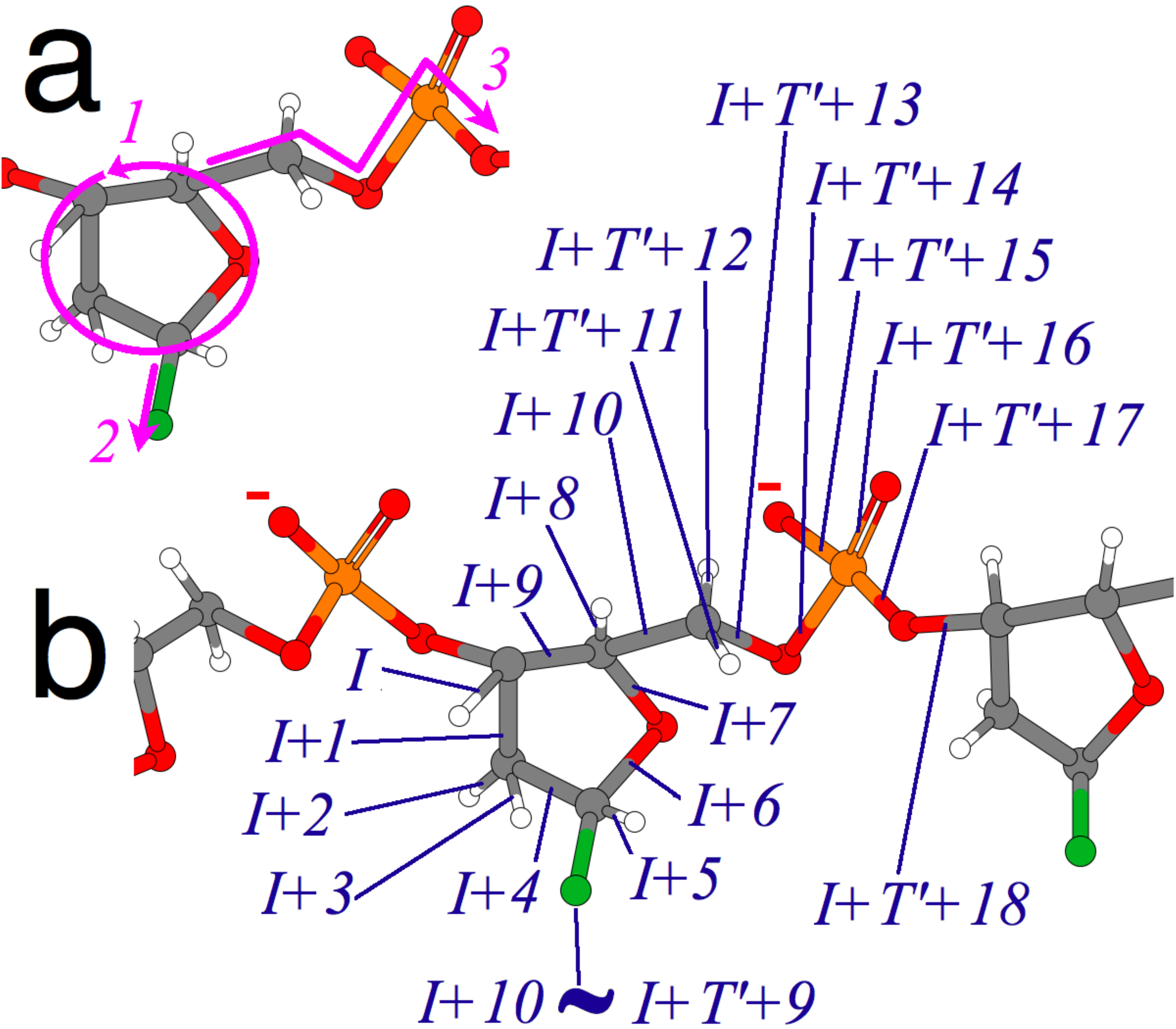}
\caption{Constraints indexing of a DNA nucleotide \textbf{a)} General order to
be followed. \textbf{b)} Indexing of the bond length constraints; $I$ denotes
the index of the first constraint imposed on the nucleotide and $T^\prime$ is
the variable number of constraints imposed on the bonds in the base.}
\label{fig:DNA_constrs}
\end{figure}

In fig. \ref{fig:DNA_constrs}, and following the same ideas as in the previous
section, we propose a way to index the bond length constraints of a DNA strand
which produces a banded matrix $R$ of low bandwidth. Green spheres represent
the (many-atom) bases (A, C, T or G), and the general path to be followed for
consecutive constraint indices is depicted in the upper left corner: first the
sugar ring, then the base and finally the rest of the nucleotide, before
proceeding to the next one in the chain.

This ordering translates into the following form for the block of $R$
corresponding to one single nucleotide dettached from the rest of the chain:
\begin{equation}
\label{RpartDNA}
R_\mathcal{M} =
\left(
\begin{array}{ccc|ccc|ccc}
 & & & & & & & & \\
 & R_\mathcal{M}^{1,1} & & & R_\mathcal{M}^{1,2} & & & R_\mathcal{M}^{1,3} & 
 \\
 & & & & & & & & \\ \hline
 & & & & & & & & \\
 & R_\mathcal{M}^{2,1} & & & S & & & & 
 \\
 & & & & & & & & \\ \hline
 & & & & & & & & \\
 & R_\mathcal{M}^{3,1} & & & & & & R_\mathcal{M}^{3,3} & 
 \\
 & & & & & & & & \\
\end{array}
\right)	\ ,
\end{equation}
where $S$ is the block associated to the constraints imposed on the bonds that
are contained in the base, $R_\mathcal{M}^{1,2}$, $R_\mathcal{M}^{1,3}$,
$R_\mathcal{M}^{2,1}$, and $R_\mathcal{M}^{3,1}$ are very sparse rectangular
blocks with only a few non-zero entries in them, and the form of the diagonal
blocks associated to the sugar ring and backbone constraints is the following:
\begin{subequations}
\label{eq:R11R33}
\begin{align}
R_\mathcal{M}^{1,1} & =
\left(
\begin{array}{cccccccccc}
\Omega & \omega & & & &  & &  & & \omega \\
\omega & \Omega & \omega & \omega & \omega & & & & & \\
 & \omega & \Omega & \omega & \omega & & & & & \\
 & \omega & \omega & \Omega & \omega & & & & & \\ 
 & \omega & \omega & \omega & \Omega & \omega & \omega & & & \\
 & & & & \omega & \Omega & \omega & & & \\
 & & & & \omega & \omega & \Omega & \omega & & \\
 & & & & & & \omega & \Omega & \omega & \omega \\
 & & & & & & & \omega & \Omega & \omega \\
\omega & & & & & & & \omega & \omega & \Omega \\
\end{array}
\right)	\ , \label{eq:R11} \\
R_\mathcal{M}^{1,1} & =
\left(
\begin{array}{ccccccccc}
\Omega & \omega & \omega & \omega & & & & & \\
\omega & \Omega & \omega & \omega & & & & & \\
\omega & \omega & \Omega & \omega & & & & & \\
\omega & \omega & \omega & \Omega & \omega & & & & \\
 & & & \omega & \Omega & \omega & \omega & \omega & \\
 & & & & \omega & \Omega & \omega & \omega & \\
 & & & & \omega & \omega & \Omega & \omega & \\
 & & & & \omega & \omega & \omega & \Omega & \omega \\
 & & & & & & & \omega & \Omega \\
\end{array}
\right)	\ . \label{eq:R33}
\end{align}
\end{subequations}

\begin{figure}
\centering
\includegraphics[width=7cm]{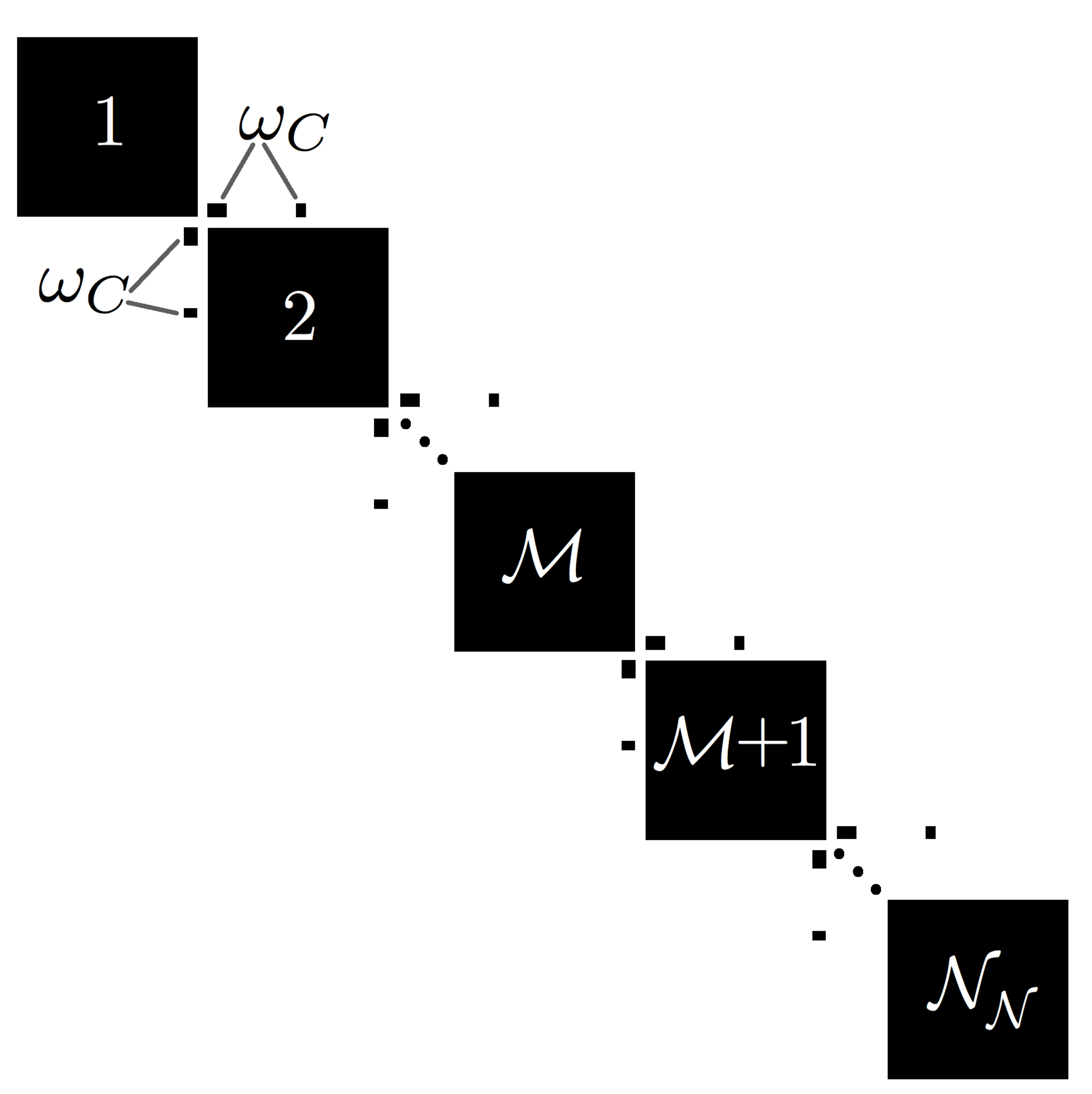}
\caption{Scheme of the matrix $R$ for a DNA molecule with $\mathcal{N_N}$ 
nucleotides. In black, we represent the potentially non-zero entries, and each 
large block in the diagonal is given by~(\ref{RpartDNA}).}
\label{fig:matrixadn}
\end{figure}

Analagously to the case of proteins, as many blocks as those in
eq.~(\ref{RpartDNA}) as nucleotides contains a given DNA strand have to be
joined to produce the global matrix $R$ of the whole molecule, together with
the $\omega_C$ blocks associated to the constraints on the bonds that connect
the different monomers. In fig.~\ref{fig:matrixadn}, a scheme of this global
matrix is depicted and we can appreciate that it indeed banded. The
construction of the matrix $R$ for a RNA molecule should follow the same steps
and the result will be very similar.

\section{Numerical calculations}
\label{sec:numerical}

In this section, we apply the efficient technique introduced in this work to a
series of polyalanine molecules in order to calculate the Lagrange multipliers
when bond length constraints are imposed. We also compare our method, both in
terms of accuracy and numerical efficiency, to the traditional inversion of 
the matrix $R$ without taking into account its banded structure.

We used the code Avogadro \cite{Avogadro} to build polyalanine chains of
$N_\mathrm{res}=$2, 5, 12, 20, 30, 40, 50, 60, 80, 90 and 100 residues, and we
chose their initial conformation to be approximately an alpha helix, i.e.,
with the values of the Ramachandran angles in the backbone $\phi=-60^o$ and
$\psi=-40^o$ \cite{Ech2007COP}. Next, for each of these chains, we used the
molecular dynamics package AMBER \cite{Cas2006UM} to produce the atoms
positions ($x$), velocities ($v$) and external forces ($F$) needed to
calculate the Lagrange multipliers (see sec.~\ref{sec_aclm}) after a short
equilibration molecular dynamics simulations. We chose to constrain all bond
lengths, but our method is equally valid for any other choice, as the more
common constraining only of bonds that involve hydrogens.

In order to produce reasonable final conformations, we repeated the following
process for each of the chains:

\begin{itemize}

\item Solvation with explicit water molecules.

\item Minimization of the solvent positions holding the polypeptide chain 
fixed (3,000 steps).

\item Minimization of all atoms positions (3,000 steps).

\item Thermalization: changing the temperature from 0 K to 300 K during 10,000 
molecular dynamics steps.

\item Stabilization: 20,000 molecular dynamics steps at a constant temperature
of 300 K.

\item Measurement of $x$, $v$ and $F$.
\end{itemize}

Neutralization is not necessary, because our polyalanine chains are themselves
neutral. In all calculations we used the force field described in
\cite{Dua2003JCC}, chose a cutoff for Coulomb interactions of 10 \AA{ } and a
time step equal to 0.002 ps, and impose constraints on all bond lengths as
mentioned. In the thermostated steps, we used Langevin dynamics with a
collision frequency of 1 ps$^{-1}$.

\begin{figure}[!b]
\begin{center}
\includegraphics[width=8cm]{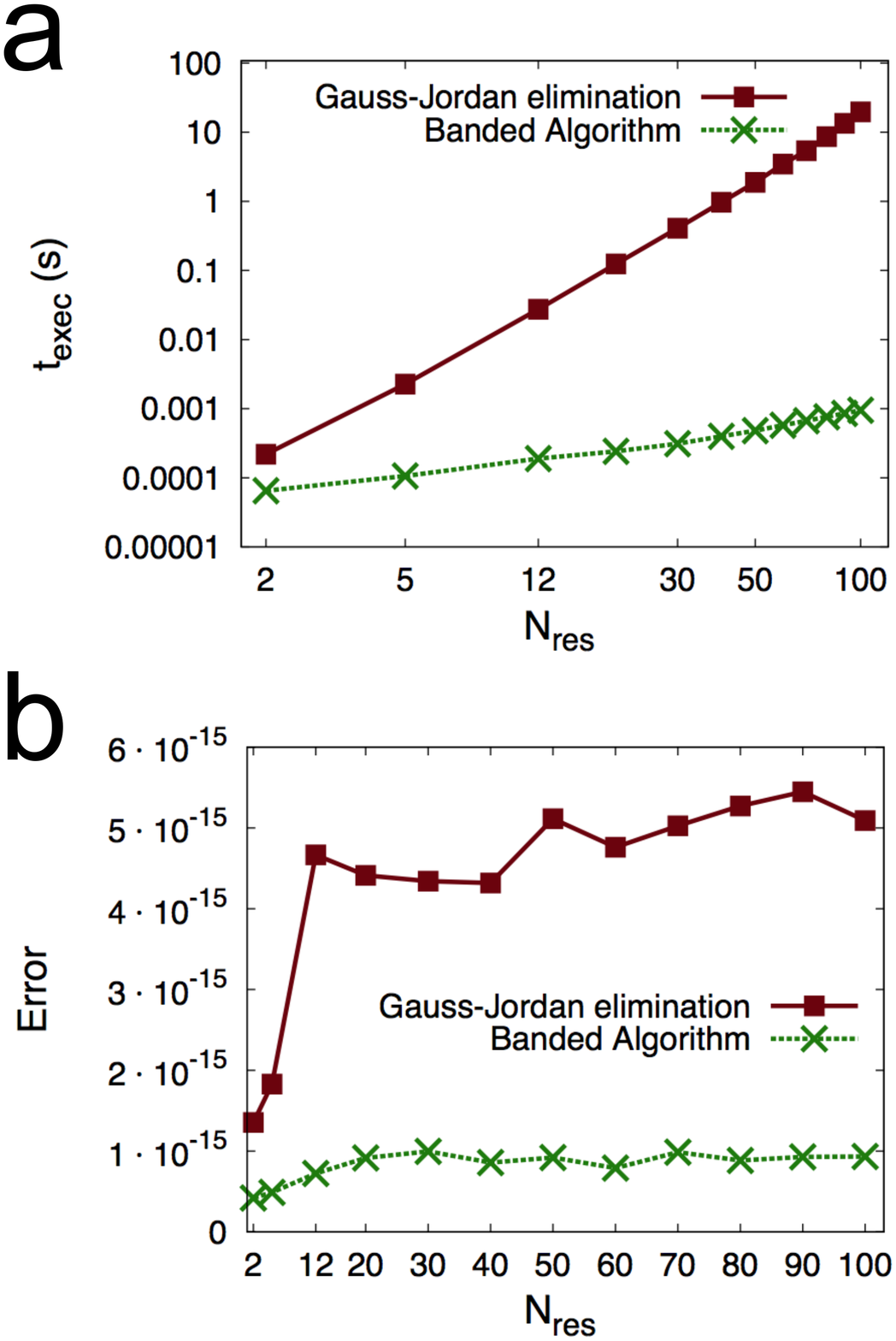}
\caption{Comparison of \textbf{a)} numerical complexity and \textbf{b)}
accuracy between a traditional Gauss-Jordan solver (solid line) and the banded
algorithm described in this work (dashed line), for the calculation of the Lagrange
multipliers on a series of polyalanine chains as a function of their number of
residues $N_\mathrm{res}$.}
\label{fig:pruebas1}
\end{center}
\end{figure} 

Using the information obtained and the indexing of the constraints described
in this work, we constructed the matrix $R$ and the vector $o$ and proceeded
to find the Lagrange multipliers using eq.~(\ref{lm}). Since~(\ref{lm}) is a
linear problem, one straightforward way to solve is to use traditional
Gauss-Jordan elimination or LU factorization \cite{Gol2002BOOK,Pre2007BOOK}.
But these methods have a drawback: they scale with the cube of the size of the
system. I.e., if we imposed $N_c$ constraints on our system (and therefore we
needed to obtain $N_c$ Lagrange multipliers), the number of floating point
operations that these methods would require is proportional to $N_{c}^{3}$.
However, as we showed in the previous sections, the fact that many biological
molecules, and proteins in particular, are essentially linear, allows to
index the constraints in such a way that the matrix $R$ in eq.~(\ref{lm})
is banded and use different techniques for solving the problem which
require only $O(N_c)$ floating point operations \cite{GR2010JCoP}.

\begin{table}\label{table:pruebas}
\begin{center}
\begin{tabular}{|c | c|c | c|c|}
\hline
$N_\mathrm{res}$ & Gauss-Jordan & Banded Alg. & Gauss-Jordan & Banded Alg.  \\ 
 & Error & Error & $t_\mathrm{exec}$ (s)  &  $t_\mathrm{exec}$ (s) \\ \hline
 2 & $1.355\cdot 10^{-15}$ & $ 4.193 \cdot 10^{-16}$ & $2.185 \cdot 10^{-4}$ & $6.500 \cdot 10^{-5}$ \\ \hline 
5 & $ 1.829\cdot 10^{-15}$ & $ 4.897\cdot 10^{-16}$ & $2.263 \cdot 10^{-3}$ & $1.059 \cdot 10^{-4}$ \\ \hline
 12 & $ 4.660 \cdot 10^{-15}$ & $ 7.244\cdot 10^{-16}$ & $2.733 \cdot 10^{-2}$ & $1.897  \cdot 10^{-4}$ \\ \hline
 20    &  $4.413\cdot 10^{-15}$ & $ 9.160\cdot 10^{-16}$  &  0.1239 &  $2.407 \cdot 10^{-4}$ \\ \hline
 30    &  $ 4.340\cdot 10^{-15}$ &  $ 9.975\cdot 10^{-16}$ &  0.4075 &   $3.115 \cdot 10^{-4}$ \\ \hline
 40    &  $ 4.318\cdot 10^{-15}$ &  $ 8.591\cdot 10^{-16}$ &  0.9669 &  $3.975 \cdot 10^{-4}$ \\ \hline
 50    &  $ 5.113\cdot 10^{-15}$ & $ 9.209\cdot 10^{-16}$  &  1.877 &  $4.811 \cdot 10^{-4}$ \\ \hline
 60    &   $ 4.761\cdot 10^{-15}$ & $ 7.906\cdot 10^{-16}$  & 3.457  & $5.751 \cdot 10^{-4}$ \\ \hline
 70    &  $ 5.026\cdot 10^{-15}$ &  $ 9.868\cdot 10^{-16}$ &  5.381 & $6.664 \cdot 10^{-4}$ \\  \hline
 80    &  $ 5.271 \cdot 10^{-15}$ & $ 8.843\cdot 10^{-16}$  & 8.633  & $7.605 \cdot 10^{-4}$ \\ \hline
  90     & $ 5.448\cdot 10^{-15}$ & $ 9.287\cdot 10^{-16}$ & 13.42 & $8.527 \cdot 10^{-4}$ \\ \hline
  100     & $ 5.091 \cdot 10^{-15}$ & $ 9.342\cdot 10^{-16}$ & 19.69 &  $9.484 \cdot 10^{-4}$ \\ \hline
\end{tabular} 
\caption{Comparison of numerical complexity and accuracy between a traditional
Gauss-Jordan solver and the banded algorithm described in this work, for the
calculation of the Lagrange multipliers on a series of polyalanine chains as a
function of their number of residues $N_\mathrm{res}$.}
\end{center}
\end{table}

In fig.~\ref{fig:pruebas1} and table \ref{table:pruebas}, we compare both the
accuracy and the execution time of the two different methods: Gauss-Jordan
elimination \cite{Pre2007BOOK}, and the banded recursive solution advocated
here and made possible by the appropriate indexing of the constraints. The
calculations have been run on a Mac OS X laptop with a 2.26 GHz Intel Core 2
Duo processor, and the errors were measured using the normalized deviation of
$R \lambda$ from $-o$. I.e., if we denote by $\lambda$ the solution provided
by the numerical method,
\begin{equation}
\label{Err}\mathrm{Error}:=\frac{\sum_{I=1}^{N_c}\left|\sum_{J=1}^{N_c}
        R_{IJ}\lambda_{J}+o_{I}\right|}
       {\sum_{I=1}^{N_c}|\lambda_{I}|} \ .
\end{equation}

From the obtained results, we can see that both methods produce an error which
is very small (close to machine precision), being the accuracy of the banded
algorithm advocated in this work slightly higher. Regarding the computational
cost, as expected, the Gauss-Jordan method presents an effort that
approximately scales with the cube of the number of constraints $N_c$ (which
is approximately proportional to $N_\mathrm{res}$), while the banded technique
allowed by the particular structure of the matrix $R$ follows a rather
accurate lineal scaling. Although it is typical that, when two such different
behaviours meet, there exists a range of system sizes for which the method
that scales more rapidly is faster and then, at a given system size, a
crossover takes place and the slower scaling method becomes more efficient
from there on, in this case, and according to the results obtained, the banded
technique is less time-consuming for all the explored molecules, and the
crossover should exist at a very small system size (if it exists at all). This
is very relevant for any potential uses of the methods introduced in this
work.

\section{Conclusions}
\label{sec:conclusions}

We have shown that, if we are dealing with typical biological polymers, whose
covalent connectivity is that of essentially linear objects, the Lagrange
multipliers that need to be computed when $N_c$ constraints are imposed on
their internal degrees of freedom (such as bond lengths, bond angles, etc.)
can be obtained in $O(N_c)$ steps as long as the constraints are indexed in a
convenient way and banded algorithms are used to solve the associated linear
system of equations. This path has been traditionally regarded as too costly
in the literature \cite{Ryc1977JCOP,Dil1987JCC,Cic1986CPR,Kra2000JCC,Gon2006JCP,Maz2007JPA}, and,
therefore, our showing that it can be implemented efficiently could have
profound implications in the design of future molecular dynamics algorithms.

Since the field of imposition of constraints in moleculary dynamics
simulations is dominated by methods that cleverly achieve that the system
exactly stays on the constrained subspace as the simulation proceeds by not
calculating the exact Lagrange multipliers, but a modification of them instead
\cite{Ryc1977JCOP,Hes1997JCC}, we are aware that the application of the new
techniques introduced here is not a direct one. However, we are confident that
the low cost of the new method and its close relationship with the problem of
constrained dynamics could prompt many advances, some of which are already
being pursued in our group. Among the most promising lines, we can mention a
possible improvement of the SHAKE method \cite{Ryc1977JCOP} by the use of the
exact Lagrange multipliers as a guess for the iterative procedure that
constitutes its most common implementation. Also, we are studying the
possibility of solving the linear problems that appear either in a different
implementation of SHAKE (mentioned in the original work too
\cite{Ryc1977JCOP}) or in the LINCS method \cite{Hes1997JCC}, and which are
defined by matrices which are different from but related to the matrix $R$
introduced in this work, being also banded if an appropriate indexing of the
constraints is used. Finally, we are exploring an extension of the ideas
introduced here to the calculation not only of the Lagrange multipliers but
also of their time derivatives, to be used in higher order integrators than
Verlet.

\section*{Acknowledgements}

\hspace{0.5cm} We would like to thank Giovanni Ciccotti for illuminating
discussions and wise advices, and Claudio Cavasotto and Isa\'{\i}as Lans for
the help with the setting up and use of AMBER. The numerical calculations have
been performed at the BIFI supercomputing facilities; we thank all the staff
there for the help and the technical assistance.

This work has been supported by the grants FIS2009-13364-C02-01 (MICINN,
Spain), Grupo de Excelencia ``Biocomputación y F\'{\i}sica de Sistemas
Complejos'', E24/3 (Arag\'on region Government, Spain), ARAID and Ibercaja
grant for young researchers (Spain). P. G.-R. is supported by a JAE Predoc
scholarship (CSIC, Spain).

\phantomsection
\addcontentsline{toc}{section}{References}
%\bibliography{refs}

\end{document}